# Spectro-polarimetry at the Pic du Midi Turret Dome and new observations of the solar CaII K line

J.-M. Malherbe, Observatoire de Paris, PSL Research University, CNRS, LESIA, Meudon, France

Email : Jean-Marie.Malherbe@obspm.fr ORCID id : https://orcid.org/0000-0002-4180-3729

Th. Roudier, Observatoire Midi Pyrénées, Université Paul Sabatier, CNRS, IRAP, Toulouse, France

Email : Thierry.Roudier@irap.omp.eu

## ABSTRACT

We summarize in this paper the spectro-polarimetric methods used at the Pic du Midi Turret Dome in spectroscopic or imagery mode. The polarimeters and spectrograph allow the cartography of solar magnetic fields at high spatial resolution through the Zeeman effect or measurements of the unresolved turbulent magnetic fields in the quiet Sun through the Hanle effect. We describe in this paper the optical capabilities of the successive versions of the polarimeters operating since 2003, and we present new results of magnetic field analysis with the CaII K 3933.7 Å spectral line.

## INTRODUCTION

The Pic du Midi Turret Dome, or “lunette Jean Rösch” (LJR), is dedicated to solar physics and started observations in 1959 (Roudier *et al*, 2022) in imagery mode. It is a well known instrument which produced, before year 2000, images of the granulation among the best available world wide, together with the Dunn telescope at Sacramento Peak (USA), the one meter Swedish telescope at La Palma (Spain) or the Hinode space telescope (JAXA/NASA). Solar spectroscopy was introduced at Pic du Midi in 1956 with a coelostat and a 50 cm telescope, and developed during the International Geophysical Year (IGY 1957-58) in a specific laboratory. However, the image quality was not satisfying, and it was decided in the seventies to move to the Turret Dome at the estern crest of the Pic du Midi (Mouradian *et al*, 1980). A new spectrograph, adapted to the equatorial mount, was built and commissioned in the early eighties. As the refractor is almost polarization free, polarimetry was introduced in the nineties with the grid method and a birefringent beamsplitter (Semel, 1980); it developed in a new way after year 2000 with liquid crystals, allowing the measurements of strong magnetic fields via the Zeeman effect, or the scattering polarization of spectral lines and weak turbulent magnetic fields via the Hanle depolarization. We summarize in this paper progresses in spectro-polatimetry at the Turret Dome and we present still unpublished results got with the CaII K 3933.7 Å spectral line.

Section 1 describes shortly the Pic du Midi Turret Dome, while Section 2 details the liquid crystal polarimeters and their capabilities. Section 3 is dedicated to the spectrograph, while Section 4 shows what can be done in imaging polarimetry with filters. Section 5 presents the last state-of-the-art version of the polarimeters, while Section 6 delivers new results of magnetometry in the CaII K line.

## 1 – THE REFRACTOR OF THE TURRET DOME

The Pic du Midi Turret Dome or LJR is a 50 cm aperture refractor (focal length of 6.45 m at $\lambda$= 550 nm, primary focus F1) supported by a strong equatorial mount. The dome has a very small volume and is hermetically closed (figure 1); the 50 cm lens is rejected far away, outside the turbulence layer. It is located at the East of the Pic du Midi in mostly laminar air flows (probably the best place of the observatory).

The beam has an axial symmetry along the optical axis. Polarization analysis can be achieved before light transmission into the spectrograph by a flat mirror at 45°. Hence, instrumental polarization is minimized.

The spatial resolving power (including the refractor and the spectrograph) is 0.3" in the yellow part of the spectrum. The primary image at focus F1 is magnified 5 or 10 times at the secondary focus F2 where the slit of the spectrograph is located, according to the magnification lens, which provides an equivalent focal length of 30 m or 65 m, so that the spectrograph operates either at f/60 or f/130.

Either 35 mm or 70 mm films (45 m length) were used until the introduction of CCD cameras at the end of the nineties. The films were digitized for scientific analysis.

The stability of polarization measurements is an important point. It may be affected by several effects, such as instrumental variations. It was noticed that the image quality deteriorates when the objective holder is heated by the sun. For that reason, P. Mehltretter suggested to protect the holder by a reflective ring with clear aperture corresponding exactly to the diameter of the lens (figure 1) in order to limit mechanical constraints. We discovered that instrumental depolarization (up to –25 %) occurs without the ring two hours before and after the meridian. With the ring, we measured only small fluctuations which are mainly due to seeing fluctuations and image motion. In conclusion, instrumental polarization may exist, but it is rather limited.

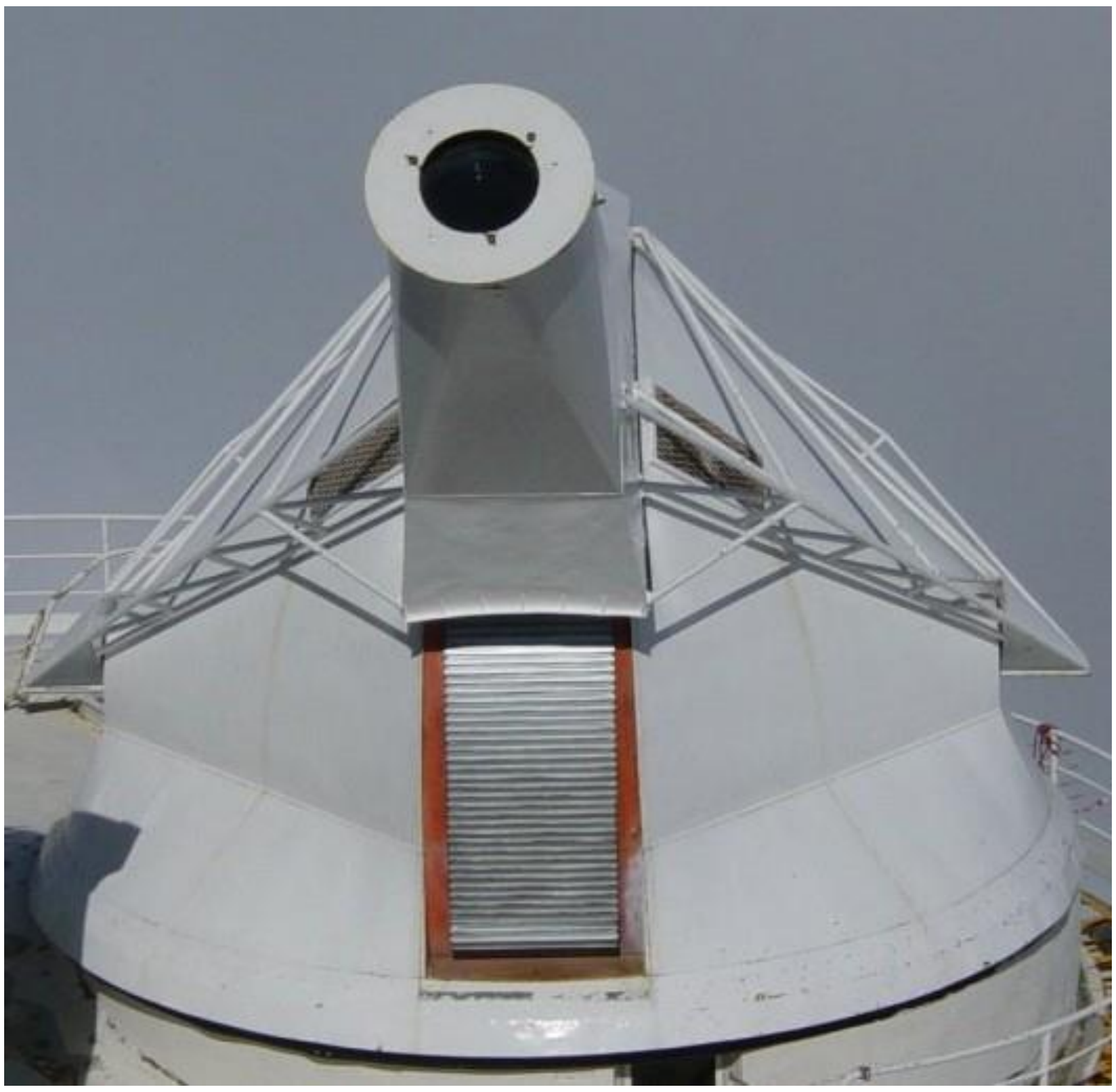

**Figure 1** : *The tube of the 50 cm/6.45 m refractor of the Pic du Midi Turret Dome (2870 m elevation). Mehltretter's ring is mounted during observations to avoid the heating of the lens holder. The dome is closed by moving curtains. Courtesy Observatoire Midi Pyrénées (OMP).*

## 2 – POLARIMETRY WITH FERRO-ELECTRIC AND NEMATIC LIQUID CRYSTALS

### 2 – a – Polarimetry with ferro-electric liquid crystals (FLC)

We used a new prototype of liquid crystal polarimeter between focus F1 and F2 in 2003 (figure 2). It was composed of an achromatic half wave plate retarder with fast modulation between +5 V and -5 V (figure 3), providind respectively the 0 and π retardances, and a linear polarizer from Meadowlark. The modulator was used alone for linear polarization or was associated with an achromatic quarter wave plate for circular polarization analysis. Hence, the output signals were:

S = ½ (I ± Q) with the modulator alone, or

S = ½ (I ± V) with the modulator and a static quarter wave plate,

where I, Q, U and V are the Stokes parameters. The modulator was protected by a bandpass filter (390-700 nm) in order to reject parasitic IR and UV light. An interference filter (100 Å FWHM) was incorporated either in polarimetric imagery of the continuum, or in spectroscopic mode for order selection by the grating. The polarimeric analysis was done between F1 and F2 along the optical axis of the refractor and was completed before light injection into the spectrograph by a flat mirror. Hence, the instrumental polarization was minimized.

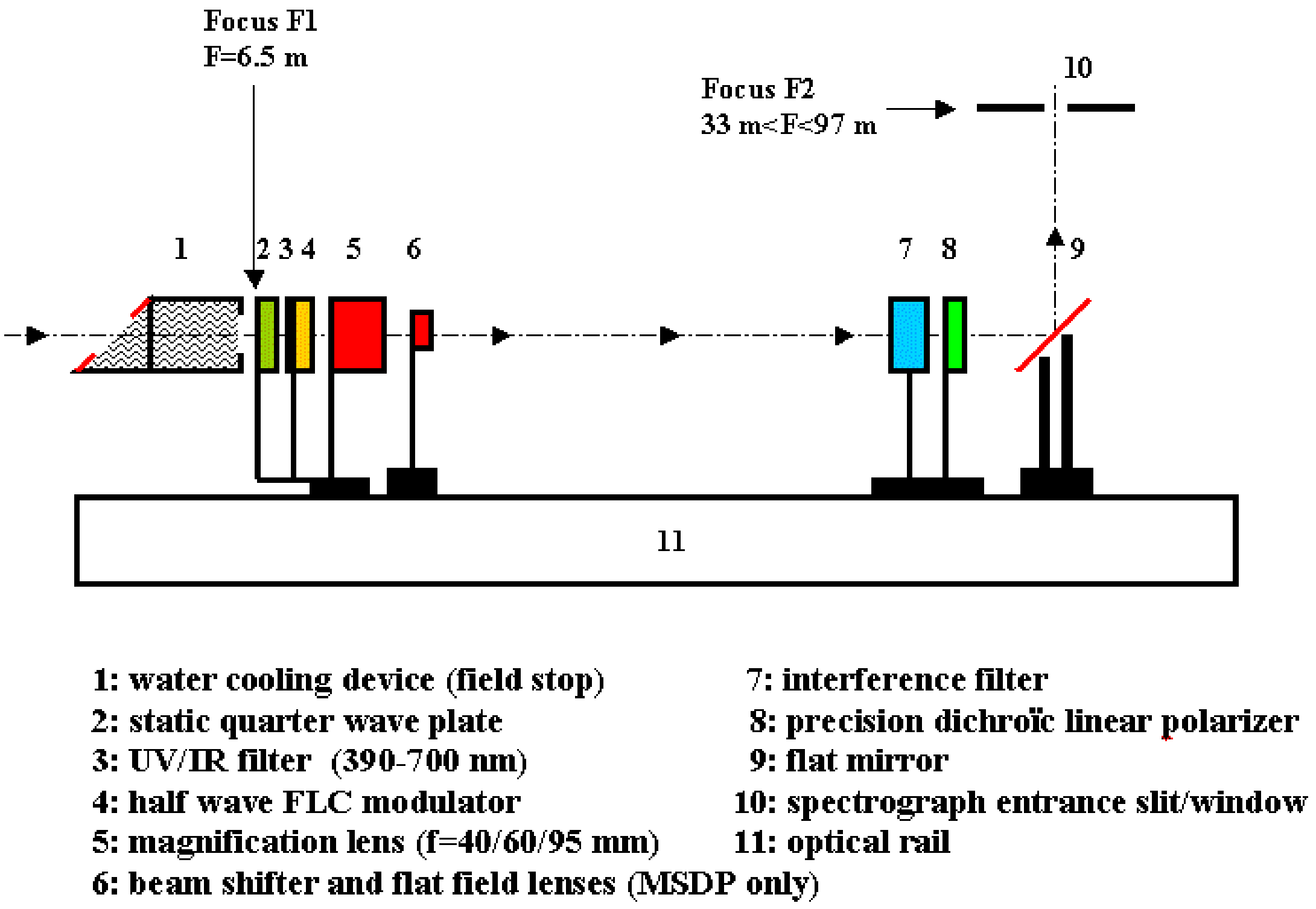

**Figure 2** : *The polarization analysis between focus F1 and F2 using a quarter wave plate (2), a half wave modulator (4) and a linear polarizer (8). Courtesy Paris observatory.*

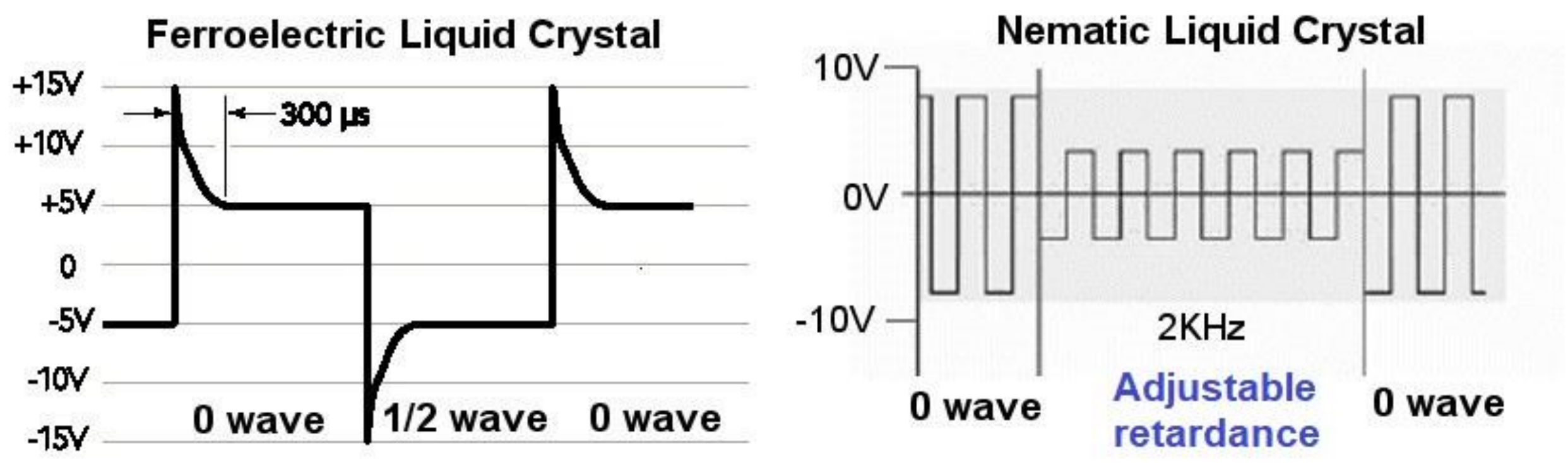

**Figure 3** : *Modulation of ferro-electric liquid cystals (FLC, left) and nematic liquid crystals (NLC, right). While FLC modulate between two fixed tensions (-5 V and +5 V, zero or half wave), NLC modulate between two adjustable values (chosen in the range -10 V to +10 V) in order to produce a variable retardance in the range zero to one wave, or 0 to 2π. FLC are faster than NLC. Courtesy Paris observatory.*

Figure 4 and 5 display observations of 19 September 2003 (Roudier *et al*, 2004) performed in imaging spectroscopy with the Multichannel Subtractive Double Pass (Mein, 1977). In this mode, the thin slit is replaced by a rectangular entrance window. It offers 11 channels presenting the same field of view, but at different wavelengths. The distance between two consecutive channels is 144 mÅ. Channels are not monochromatic, the wavelength varies linearly with the x-abscissa. Observations were performed at 5 frames/s, alternating I+V and I-V signals. The data processing is complex and allows to produce the sophisticated results of figure 5 with good temporal resolution (30 s). In particular, the continuum intensity shows the granulation (movie 1), while intensities in the line, using a 144 mÅ chord above the core, shows the low chromosphere (the granulation is no more visible) with magnetic bright points. Using the same chord, Dopplershifts are obtained by the measurement of the wavelength shift of I+V and I-V profiles, while magnetic fields are deduced from the Zeeman splitting (proportional to the field) between I+V and I-V. In order to improve the signal to noise ratio, a stacking of I+V and I-V frames (which are not simultaneous, but successively observed at fast cadence under the form of bursts of typically 100 frames) can be done after correction of the seeing effects (tip-tilt and distretching). Such sequences of observations can be repeated every 30 s in order to study the dynamics of structures evolving on short time scales (10 min), such as the granulation.

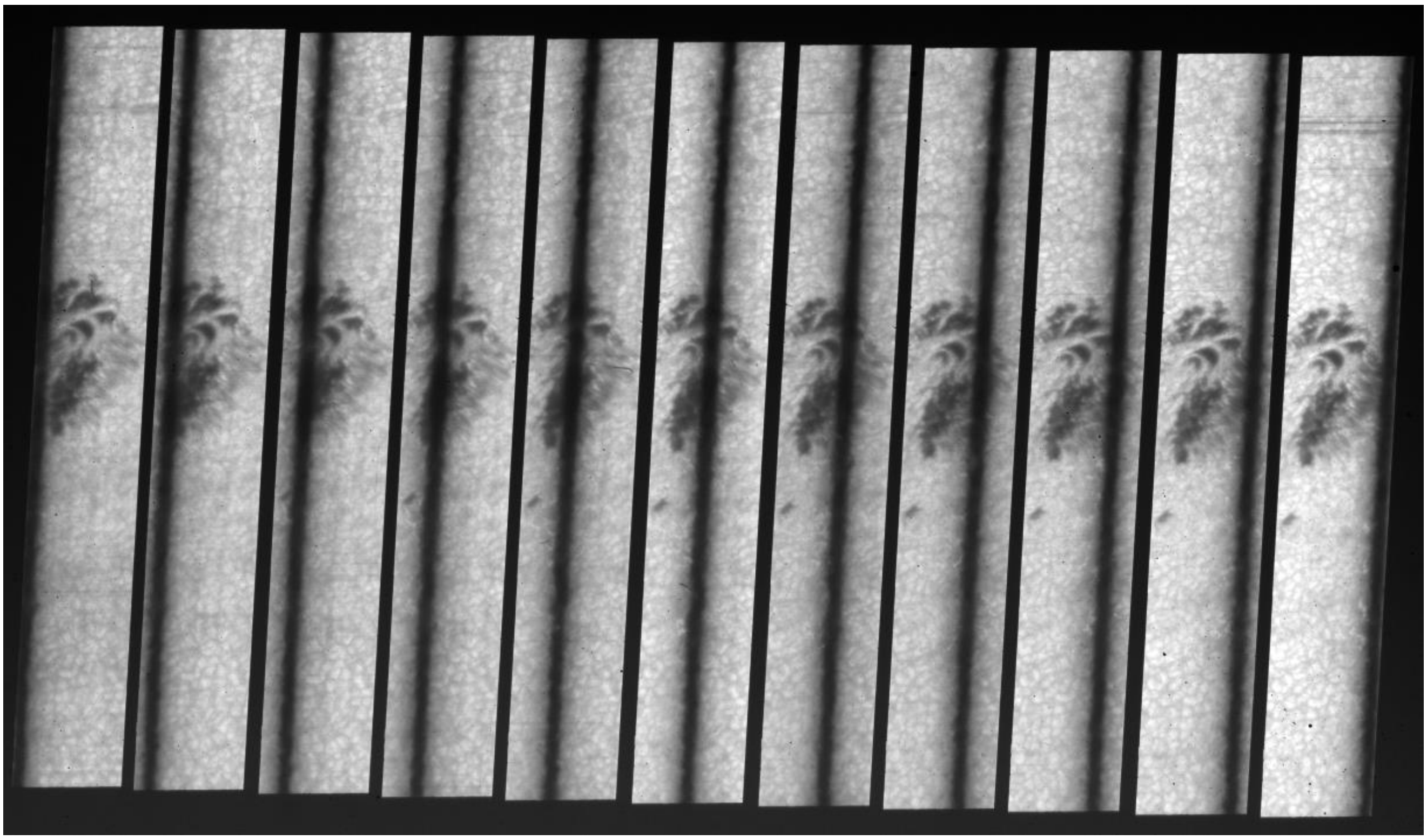

**Figure 4** : *Granulation and sunspot in MSDP imaging spectroscopy mode (NaD1 line, 11 channels). After Roudier et al (2004).*

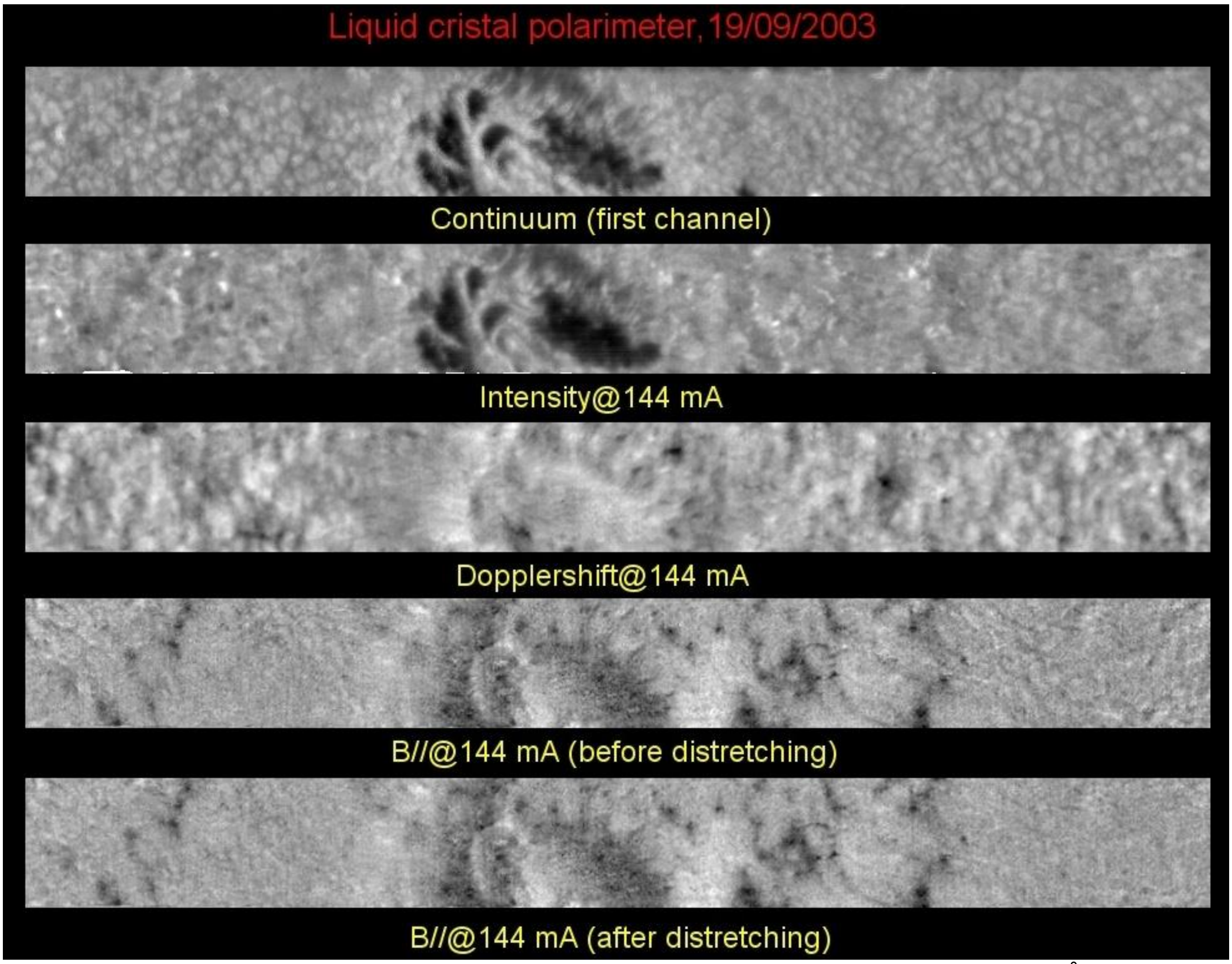

**Figure 5** : *Imaging spectroscopy in NaD1 line. Panel 1 (top): continuum; (2) intensity (144 mÅ chord); (3) Dopplershifts; (4) magnetic field; (5) magnetic field after stacking and distretching. After Roudier et al (2004).*

## 2 – b – Polarimetry with nematic liquid crystal (NLC)

The second prototype of liquid crystal polarimeter was installed between focus F1 and F2 in 2004. It was used for the first time in April 2004 in a simplified version and upgraded to the full Stokes version in September 2004. The polarimeter is composed of (figure 6) :

1) Two variable retarders (retardance adjustable between 0 and 700 nm, or zero to one wave, from Meadowlark company). A 2 kHz square wave signal (figure 3) with zero mean value is applied to the crystal; the amplitude of the signal determines the retardance. The maximum retardance is obtained for an amplitude of 0 V and decreases continuously with increasing voltage (no retardance at about 6 V). Consequently, a variable retarder can be exactly quarter or half wave for any wavelength according to the voltage applied to the electrodes. The calibration was performed in laboratory using a white light source, various interference filters of 10 nm bandwidth. It takes about 25 ms for the retarder to switch between two states using fast transient nematic effect: in order to speed up the response of the crystal, a high to low voltage transition is preceded by a short (5 ms) 0 V pulse, while a low to high voltage transition is preceded by a 10 V pulse. The NLC is protected by a passband filter in order to avoid damage by the UV and IR radiation. The temperature of the two liquid crystals is controlled permanently (active heating by resistive elements and passive cooling) and is ordinary set to 20° C.

2) A precision dichroïc linear polarizer from Meadowlark is located at the exit of the system, on the optical axis, before the reflecting mirror to the spectrograph. The acceptance axis is orthogonal to the long direction of the entrance slit in order to be orthogonal to the rules of the Echelle grating of the spectrograph (condition of maximum luminosity). The fast axis of the first retarder is parallel to the acceptance axis of the polarizer, while the fast axis of the second retarder makes an angle of 45°.

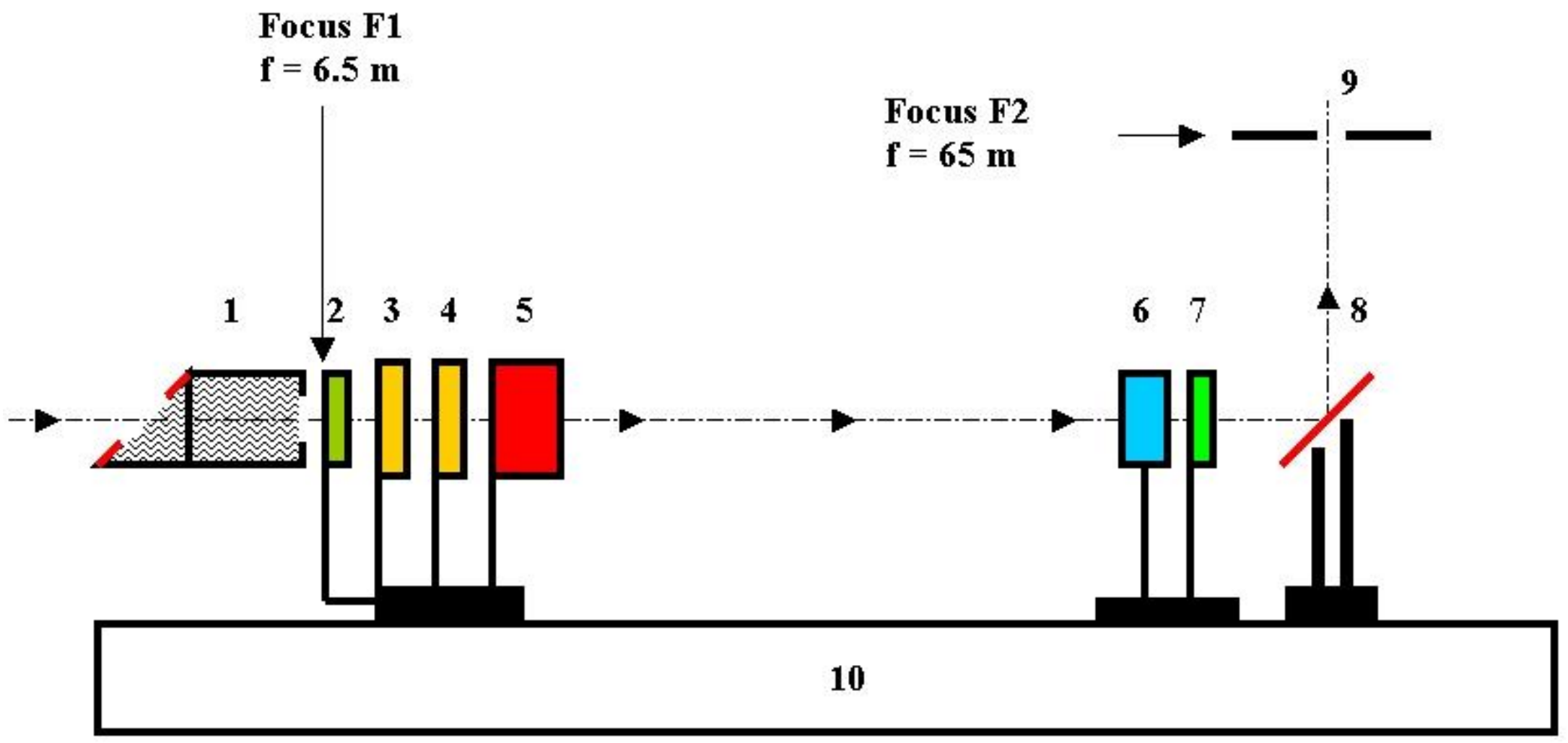

1: water cooling device (field stop)
2: UV/IR filter 380-700 nm
3: variable retarder 1
4: variable retarder 2
5: magnification lens (f=60 mm)
6: interference filter
7: precision dichroïc linear polarizer
8: flat mirror
9: spectrograph entrance slit
10: optical rail

**Figure 6** : *The polarization analysis between focus F1 and F2 using two nematic liquid crystals (3 and 4) and a linear polarizer (7). Courtesy Paris observatory.*

### Version of the polarimeter with two retarders (figure 7)

The signal S provided by the polarimeter with two NLC is given by :

$S = (1/2) ( I + Q \cos\delta 2 + \sin\delta 2 (U \sin\delta 1 – V \cos\delta 1) )$

where I, Q, U and V are the Stokes parameters. δ1, δ2 are the retardances of the two NLC. When the first retarder is set to 0 wave (δ1 = 0), and when the second retarder modulates between quarter (δ2 = π/2) and three quarter wave (δ2 = 3π/2), we get sequentially I ± V; when it modulates between 0 (δ2 = 0) and half wave (δ2 = π), we obtain I ± Q. In order to measure I ± U, the first retarder must be set to quarter wave (δ1 = π/2), while the second retarder has to modulate between quarter (δ2 = π/2) and three quarter wave (δ2 = 3 π/2). The voltage to apply is indicated by figure 8 for the two NLC as a function of wavelength.

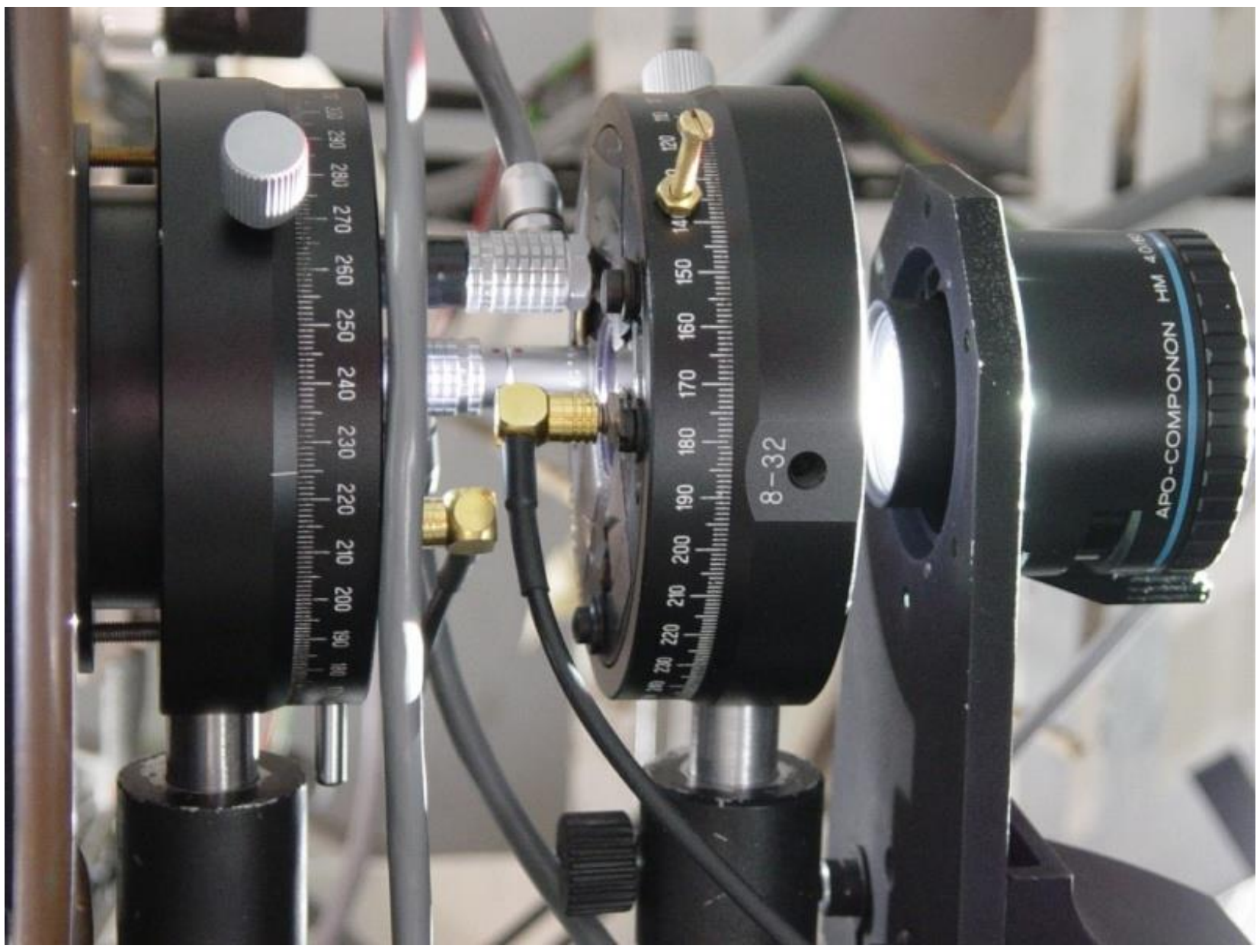

**Figure 7** : *The two variable retarders (at left) and the magnifying lens (at right) in the beam near focus F1 with wires for temperature control and modulation*

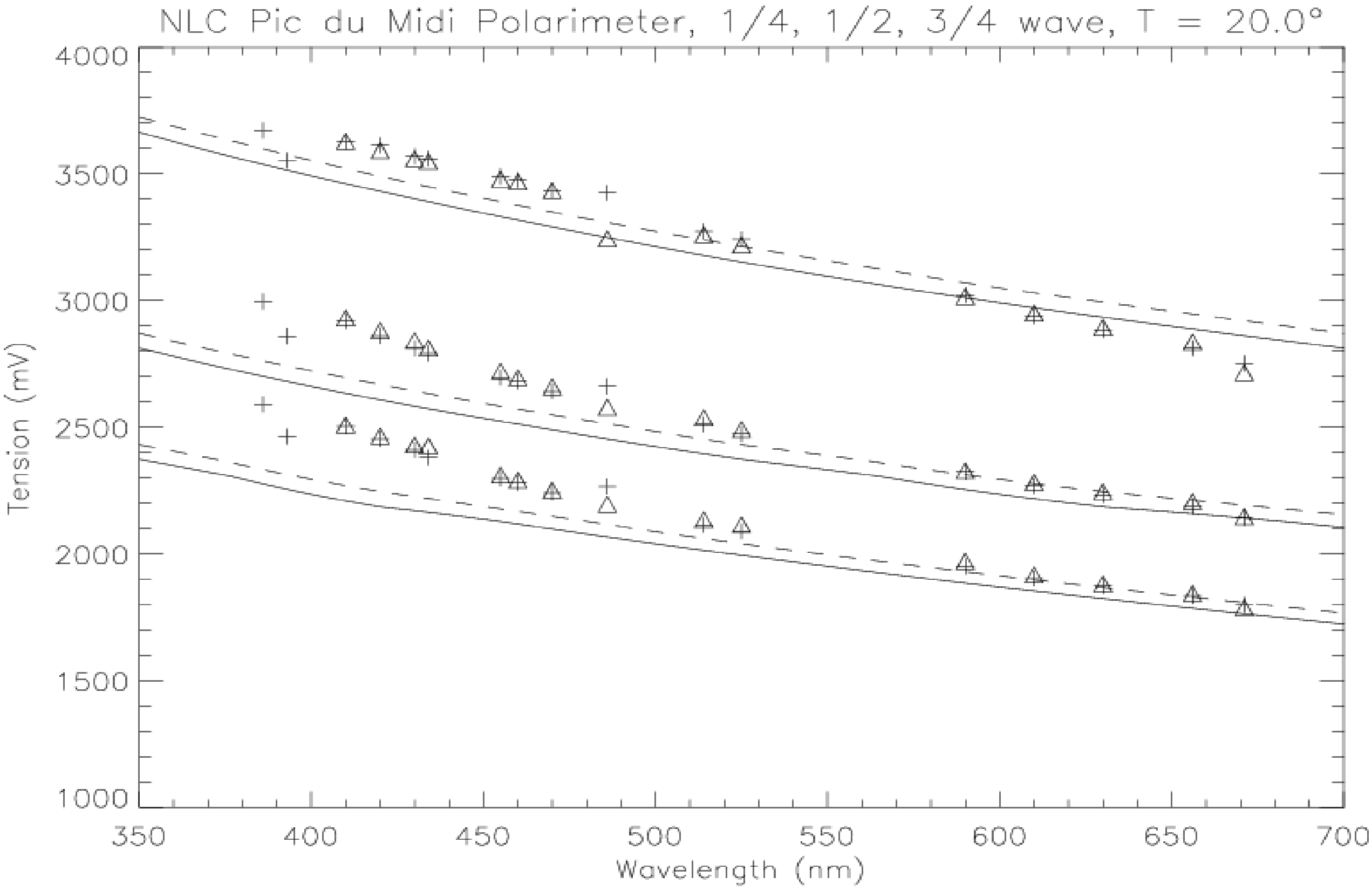

**Figure 8** : *The voltage applied to the two NLC as a function of wavelength for quarter (top), half (middle), or three quarter (bottom) wave retardance. Crosses and triangles indicate our calibration points obtained with various interference filters (10 nm bandwidth) for the two retarders at 386, 393, 410, 420, 430, 434, 455, 460, 470, 486, 514, 523, 590, 610, 630, 656 and 670 nm. For comparison, the two curves were extrapolated from data provided by the manufacturer.*

Some examples of iron spectral lines observed in full Stokes polarimetry are displayed in figure 9.

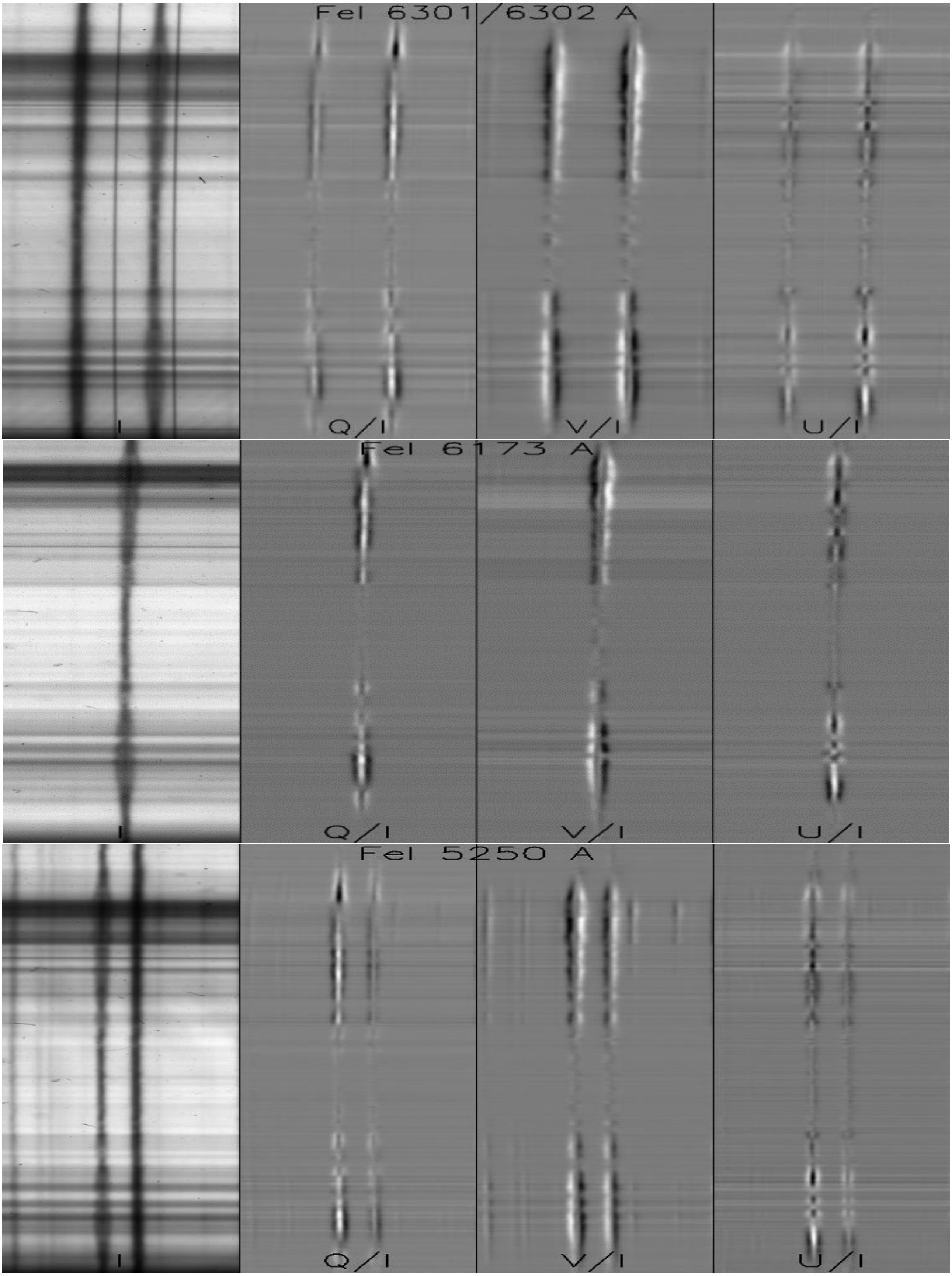

**Figure 9** : *The Zeeman effect observed in sunspots with full Stokes polarimetry in FeI 6301/6302 Å, FeI 6173 Å and FeI 5250 Å. From left to right: I and Q/I, V/I, U/I polmarization rates (abscissa: the wavelength ; ordinates: the slit direction). Courtesy Paris observatory.*

## Version of the polarimeter with one retarder

We use for observations which do not require the full Stokes polarimetry a simplified version of the polarimeter, running only with one retarder (no possible U determination). In such a configuration ($\delta 1$ is permanently set to zero), the polarimeter allows in practice the modulation from zero to three quarter waves, providing analysis either of the circular or linear polarization of light. Hence, Stokes combinations I ± V or I ± Q are obtained sequentially from the output signal :

$$S = (1/2)\,( I + Q \cos\delta - V \sin\delta)$$

with $\delta = (\pi/2, 3\pi/2)$ for I ± V and $\delta = (0, \pi)$ for I ± Q.

Since measurements of I + P and I – P (where P is any Stokes parameter) are not simultaneous, our polarimeter can operate with a good efficiency in two particular domains :

1) observations of the line of sight magnetic fields on the disk, using classical thin slit spectroscopy (figure 10) or the 2D imaging spectroscopy device provided by the Multichannel Subtractive Double Pass (MSDP), as described by Malherbe *et al* (2004) and Roudier *et al* (2006) ; within a 2D field of view, we use cross correlation and destretching methods to superimpose properly solar structures observed sequentially in two states of circular polarization from bursts of 50 couples I ± V (figure 10) ;

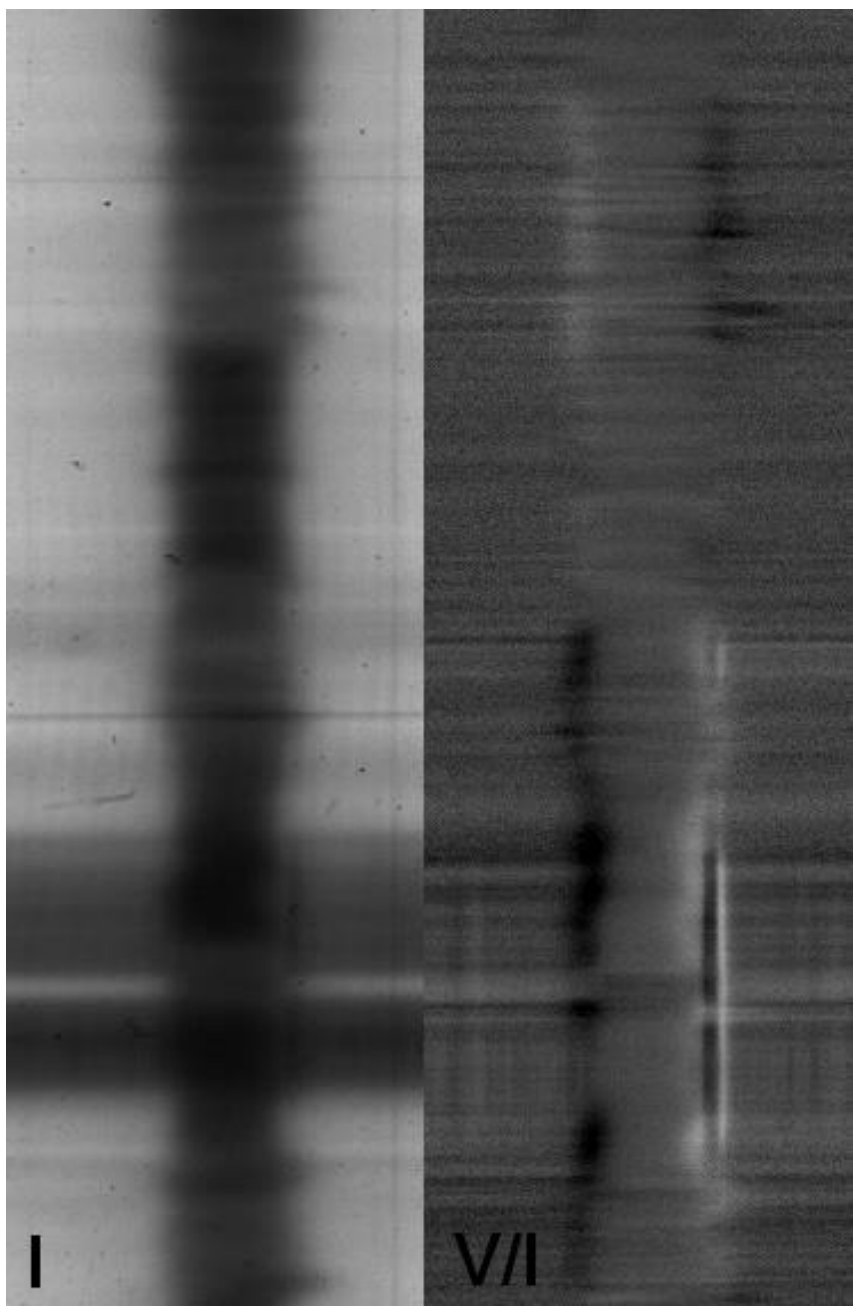

**Figure 10** : *Top: the Zeeman effect observed in sunspots and thin slit spectroscopy (circular polarization of the Hα line, I and V/I, wavelelength in abscissa). There is a blend from Cobalt in the red wing. Bottom: magnetometry in MSDP mode (imaging spectroscopy) in the NaD1 5896 Å line. The background is the solar granulation in the NaD1 continuum. Isocontours represent the line of sight magnetic field (black or white respectively for south and north polarities). Courtesy Paris observatory, after Roudier et al (2006).*

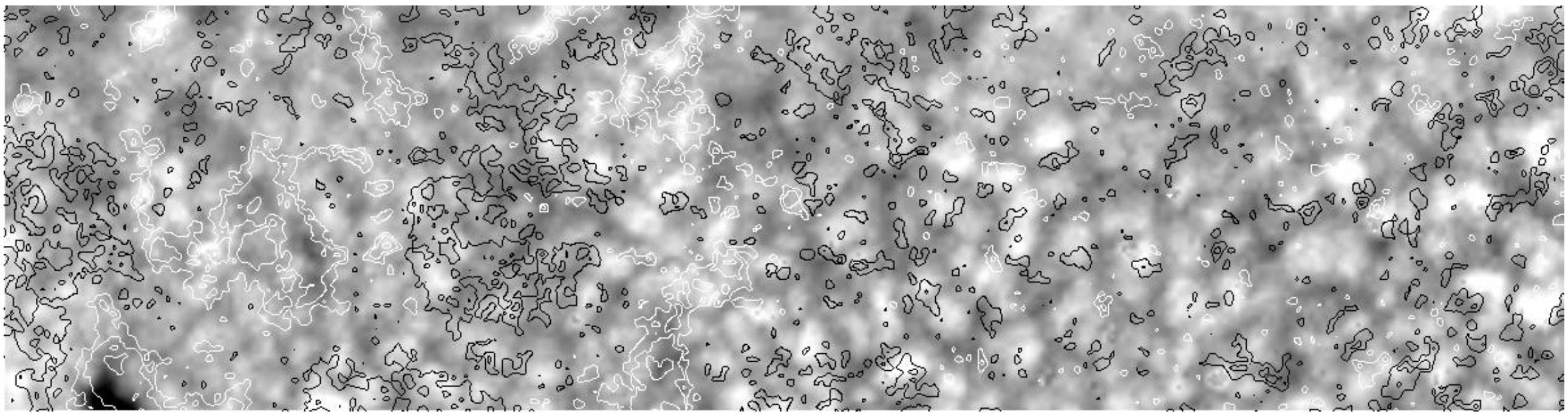

2) observations of the second solar spectrum near the limb; since the polarization rate is small (less than 1%), the signal has to be derived from the statistical analysis of hundreds or thousands of spectra obtained sequentially, in order to improve the signal to noise ratio. Consequently, non simultaneous measurements of I + Q and I – Q have a moderate impact on the final polarization rate. Since the spectrograph has a good transmission in the blue part of the spectrum, and the CCD detector has the maximum quantum efficiency (65%) around 400 - 500 nm, our observations are mainly focused in this range of the second solar spectrum. We started observations with two scientific goals:

- measurements of weak polarizations (0.01%) at moderate spatial resolution : when the seeing conditions are not excellent, we use a 0.6" x 140" slit and we accumulate a large amount of spectra to achieve the best polarimetric sensitivity as possible; data can be averaged partially or totally along the slit to reduce the noise.
- measurements of stronger polarizations (0.1% to 1%) at higher spatial resolution, as in the SrI 4607 Å line (figure 11) : when the seeing is fairly good, we use a thin slit of 0.3" x 140" in order to select structures along the slit, in terms of bright and dark regions. Such a diagnostic is able to bring new information about turbulent unresolved magnetic fields.

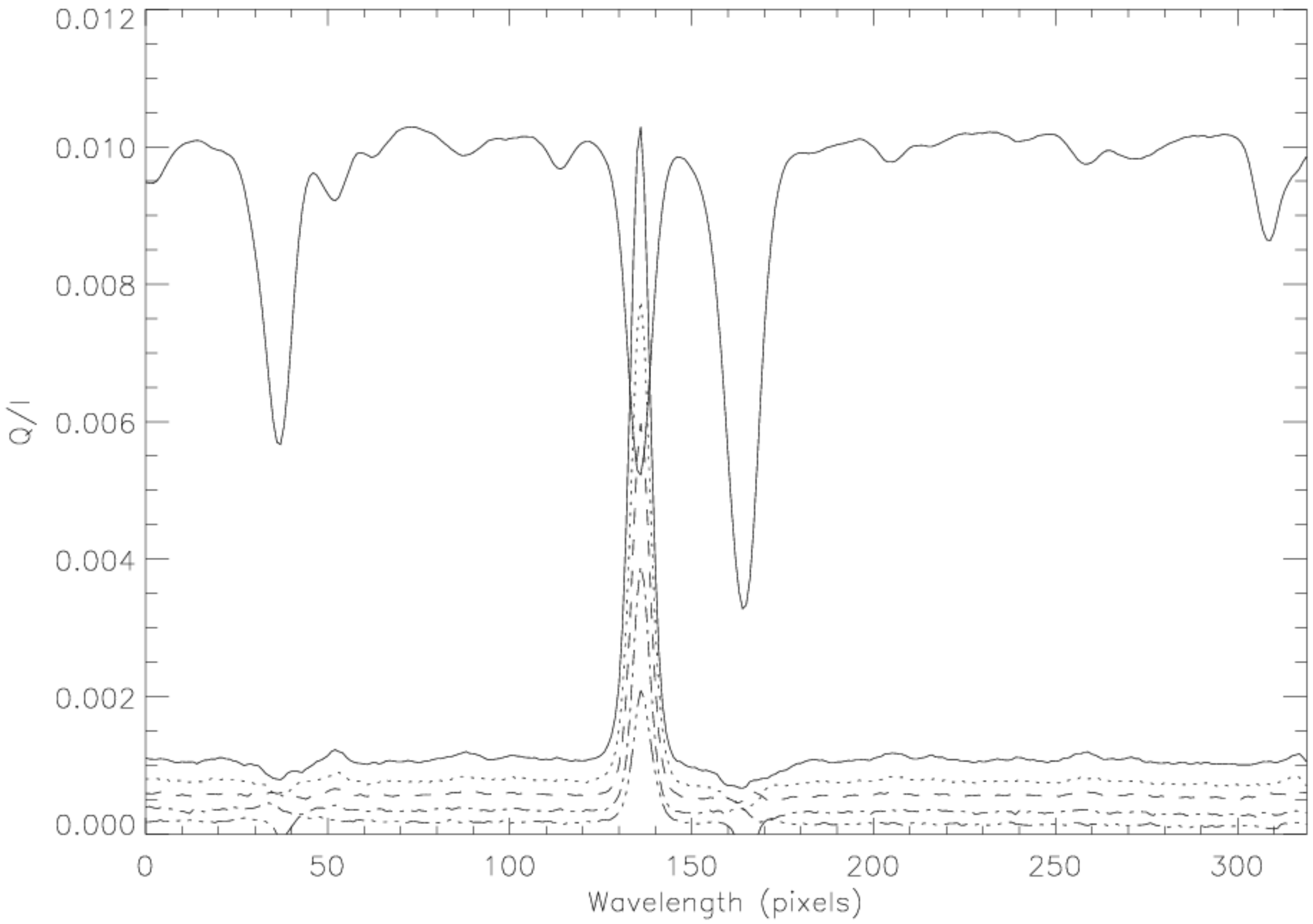

**Figure 11** : *Polarization rate (Q/I) of SrI 4607 Å line at various distances of the limb (5", 10", 20", 40", 80"). The polarization decreases with the distance, both in the line and in the continuum. Courtesy Paris observatory.*

For the second solar spectrum (figure 11), our polarimetric observations consist of shooting sequentially as fast as possible couples of images I + Q and I – Q (typically 700 pixels in the solar direction x 300 pixels in the spectral direction). The maximum speed is 10 Hz including polarimetric modulation. Since a single image I ± Q has a photometric precision of about $10^{-2}$, one hour of observation at the limb allows to reach roughly $10^{-4}$ in the blue part of the spectrum ; it is faster in the green part, but slower in the violet/UV part. But it is even possible to measure weaker polarization signals by integrating partially or totally in the solar direction along the slit (700 pixels). In the case of total integration, a single couple of images I ± Q will deliver a precision better than $10^{-3}$, and after one hour of observations, a ratio better than $10^{-5}$ can be achieved.

Flat field observations are made with telescope motions around the disk center in the quiet sun where the polarization should vanish in average. A good flat field is a difficult challenge because the spectrograph is attached to the refractor (in equatorial mount) and moves with time, producing small but permanent mechanical shifts. For that reason, the spectral line has to be tracked by the software in the direction of dispersion. For observations on the disk, the transversalium is followed along the slit direction, while for observations with the slit perpendicular to the limb, the limb position is detected by the software with about 1" accuracy, depending on seeing. Parasitic fringes do exist but cannot not be properly corrected by the flat field procedure. They are not important for polarizations in the range $10^{-3}$ to $10^{-2}$ but have to be considered carefully for polarizations of $10^{-4}$ or smaller. We have two kinds of fringes: filter fringes (mainly due to the selecting order filter) which have a level of about 1% to 2% of the continuum and vanish in the difference between two consecutive states of polarization; and polarization fringes (due to the NLC variable retarders), which are more than 100 times fainter (0.01% of the continuum) and which appear only after a long time of integration in the difference between alternate polarizations. Such fringes are hard to handle, because the residual pattern generally shifts slowly during the observing run. Flat fielding is very useful to determine precisely the transmission of the instrument for the two states of polarization (a difference of about 0.2% is typical, depending on the line, because the transmission of the NLC varies slightly with retardance).

## 3 – THE SPECTROGRAPH

We use the 8 meter Littrow Echelle spectrograph (figures 12 and 13) built by Paris Observatory fourty years ago and modified later for 2D imaging spectroscopy (MSDP). The dispersive element is a grating (316 rules/mm, blaze angle 63° 26') and provides a typical dispersion of 5 mm/Å at the focus of the spectrograph. The interference order is isolated by filters of typically 100 Å bandwidth. The spectrum obtained at the focus of the spectrograph is reduced to form on a CCD camera from LaVision (Germany) with temperature control (Peltier cooling at –10° C). This is a shutterless interline camera using microlenses manufactured by Sony (1376 x 1040, 6.45 µ pixels). The spatial pixel on the CCD is usually 0.2" along the slit direction and the spectral pixel around 10 mÅ (in the blue part of the spectrum). Each pixel can accumulate up to 20000 electrons corresponding to the dynamic of 12 bits (readout noise of 4-5 electrons, gain of 4 electrons/ADU). In general, we work in the continuum at 0.75 times the saturation level giving approximately a signal to noise ratio of 100, corresponding to the photometric accuracy of $10^{-2}$. The exposure time is typically 50 ms in the blue part of the spectrum at 10" from the limb with the 0.6" slit.

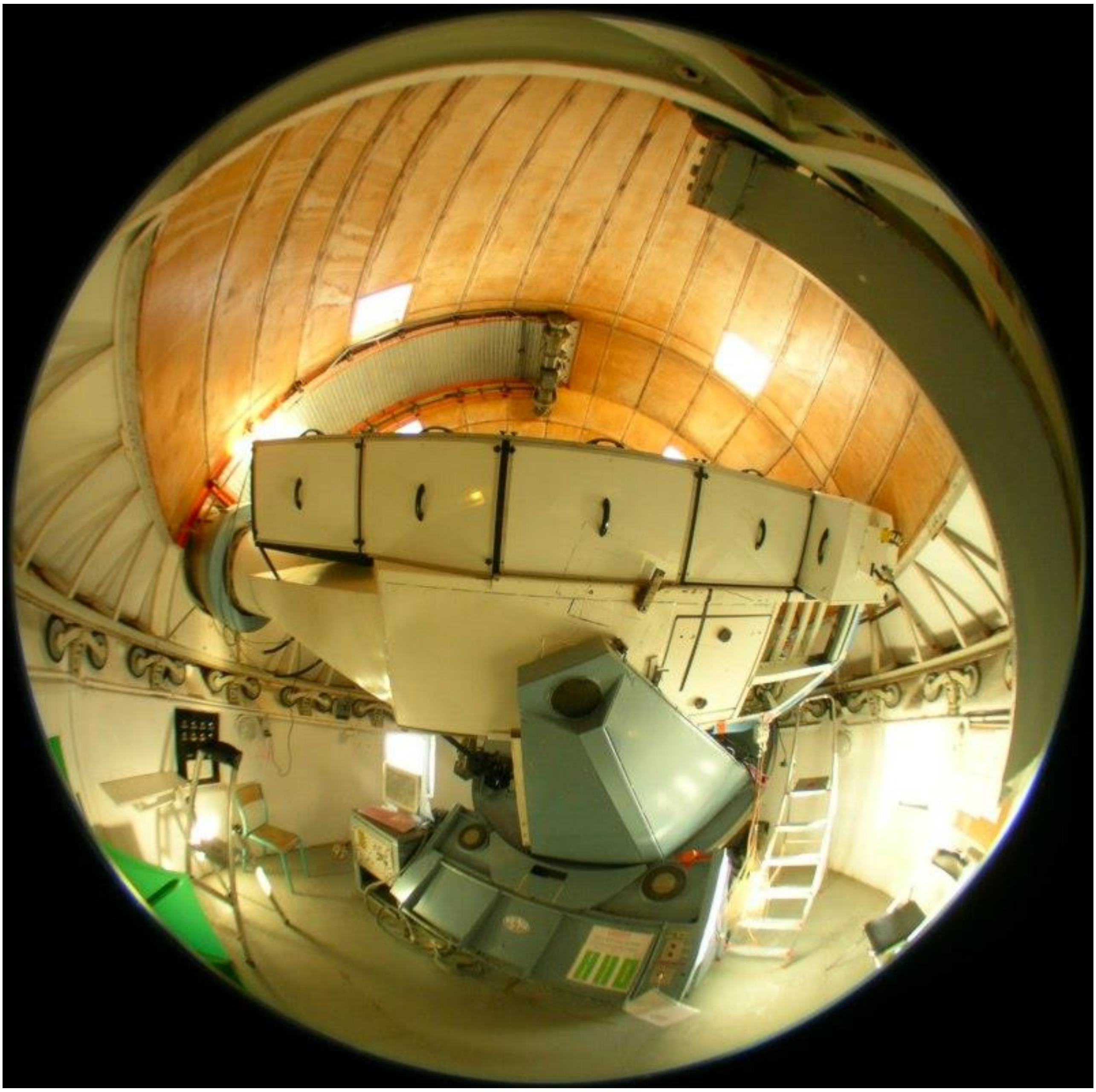

**Figure 12** : *The Turret Dome, the refractor and the spectrograph above. Courtesy Sylvain Rondi.*

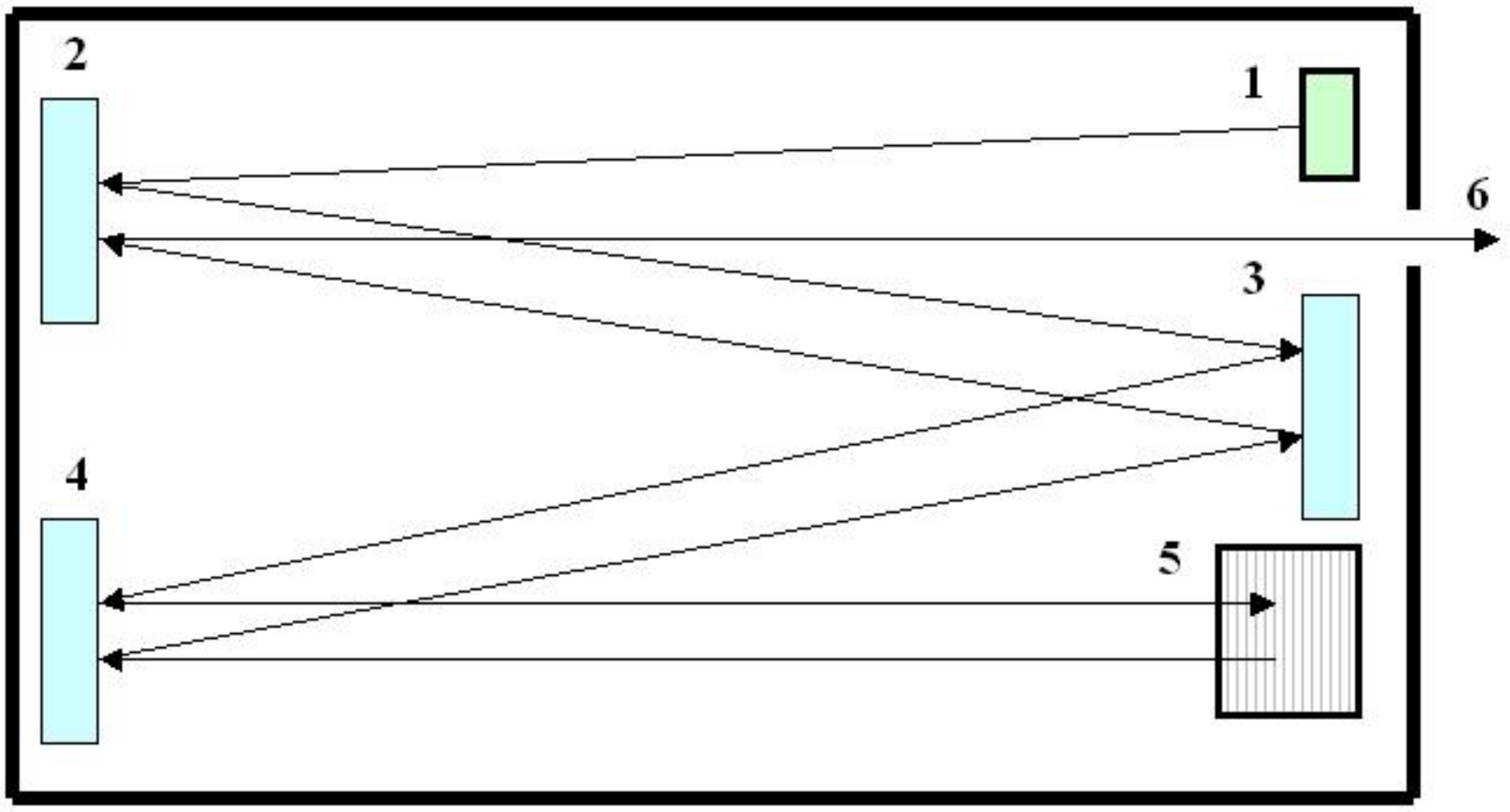

1 : flat mirror (45°)   2,3: flat mirrors   4: collimator f = 8 m
5: grating (316 groves/mm, blaze angle 63°26')
6: spectrum, transfer optics to the detector

**Figure 13** : *The 8 m focal length spectrograph of the Pic du Midi Turret Dome (the beam is folded two times in order to deal with the 3 m length of the enclosure). The light is injected into the spectrograph by mirror n°1. The transfer optics to the detector (6, not shown) is composed of a field lens (f = 1.1 m), two flat mirrors and a Nikon objective (f = 105 mm). Courtesy Paris observatory.*

## 4 – POLARIMETRY IN IMAGERY MODE

Polarimetry in imagery mode is possible. The instrumental setup is presented below (figure 14).

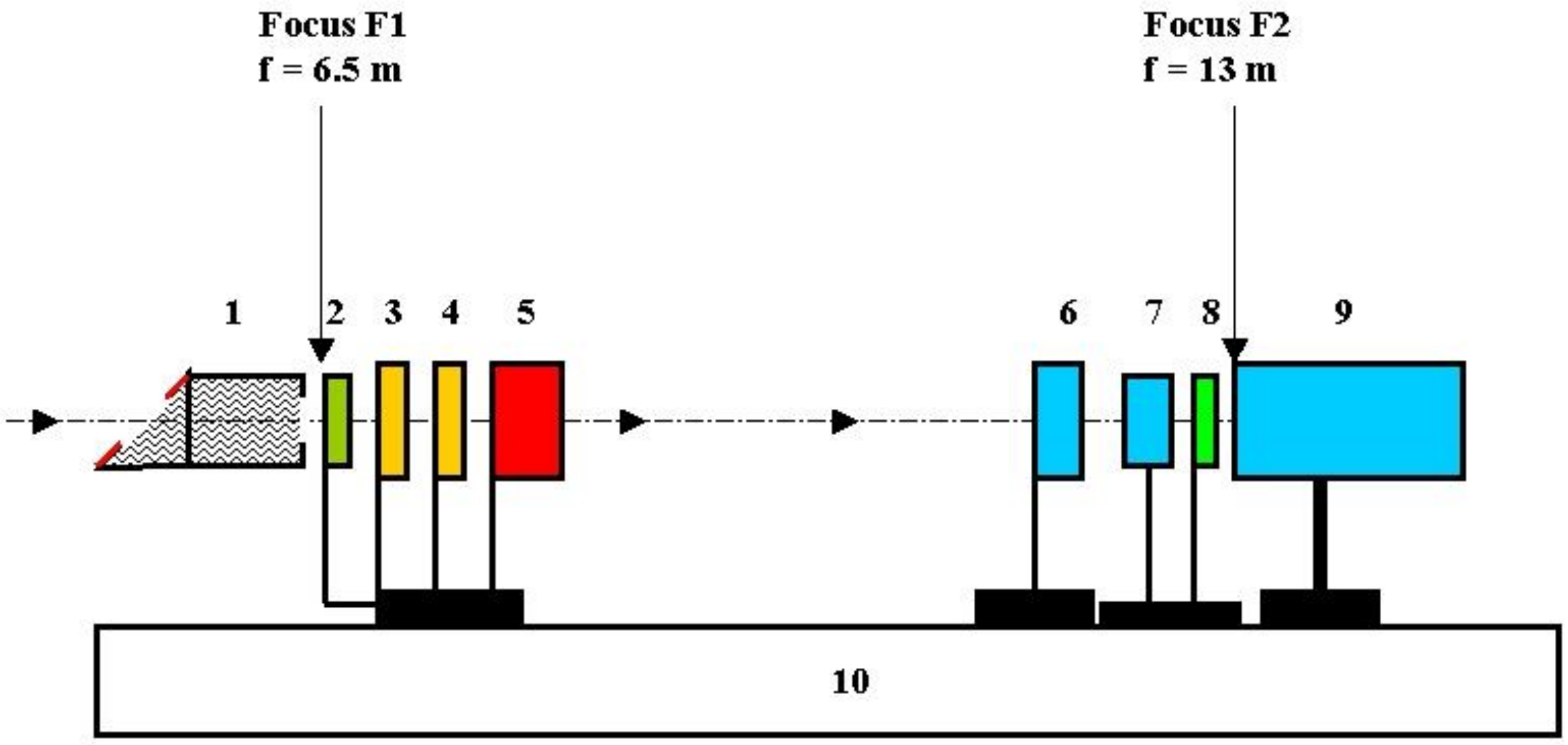

1: water cooling device (field stop)
2: UV/IR filter 380-700 nm
3: variable retarder 1
4: variable retarder 2
5: magnification lens (f=95 mm)
6: neutral density
7: interference filter
8: precision dichroïc linear polarizer
9: detector
10: optical rail

**Figure 14** : *The optical setup between the primary focus F1 and the secondary focus F2 in imagery mode (the 50 cm refractor is located on the optical axis at left). Courtesy Paris observatory.*

The polarimetric system in imagery mode provides results concerning the continuum polarization such as those of figures 15 and 16. The setup is identical to the one used in spectroscopy, but the magnification is smaller (x 2), so that we work at f/30 instead of f/65 or f/130 in spectro polarimetry. The field of view is 100" x 120" with a pixel size of 0.1". The detector is installed on the optical rail, at the location of the injection mirror to the spectrograph which is removed for the circumstance.

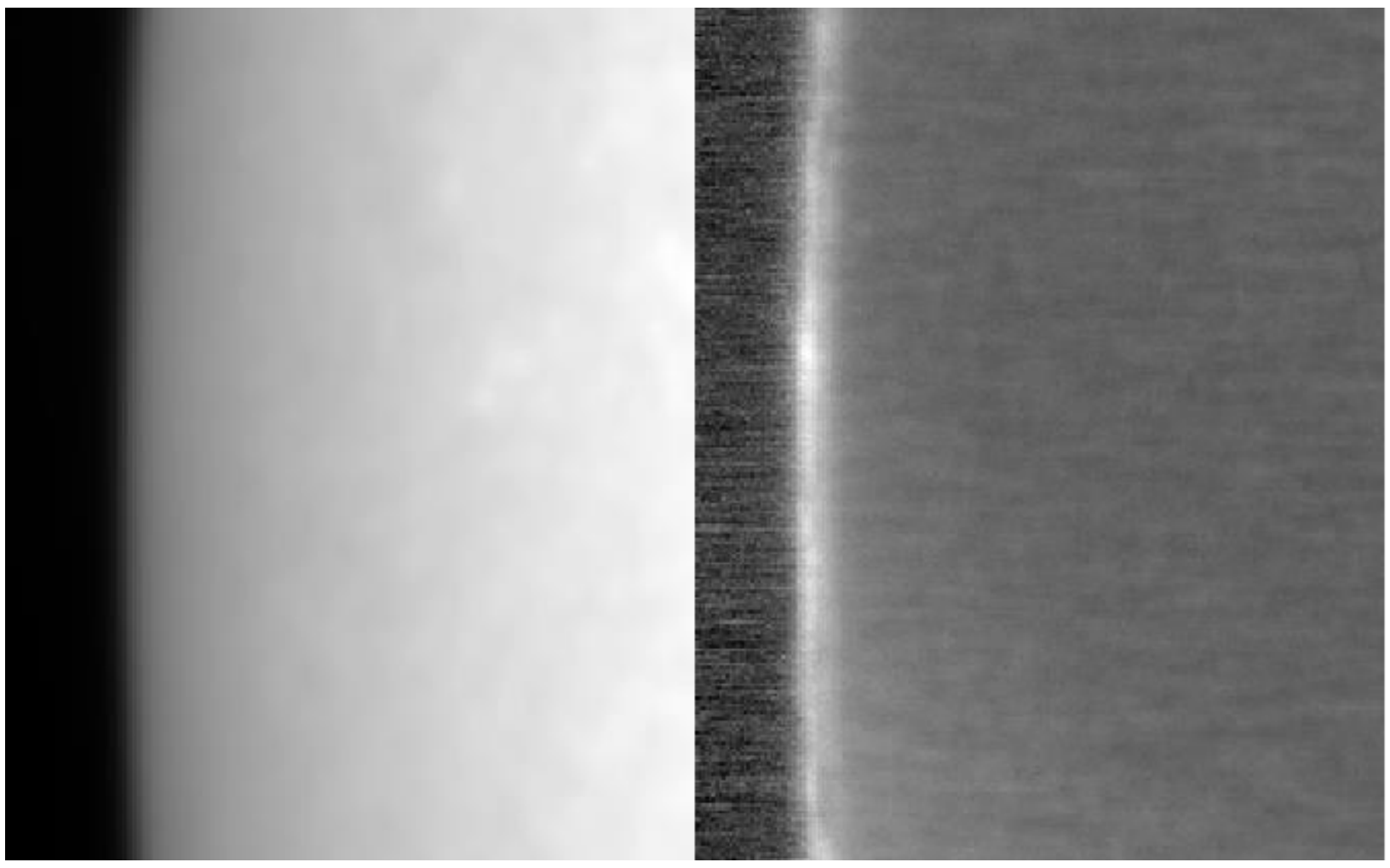

**Figure 15** : *Intensity I and linear polarization Q/I near the limb at 4600 Å (bandwidth 100 Å) in polarimetric imagery mode. Courtesy Paris observatory.*

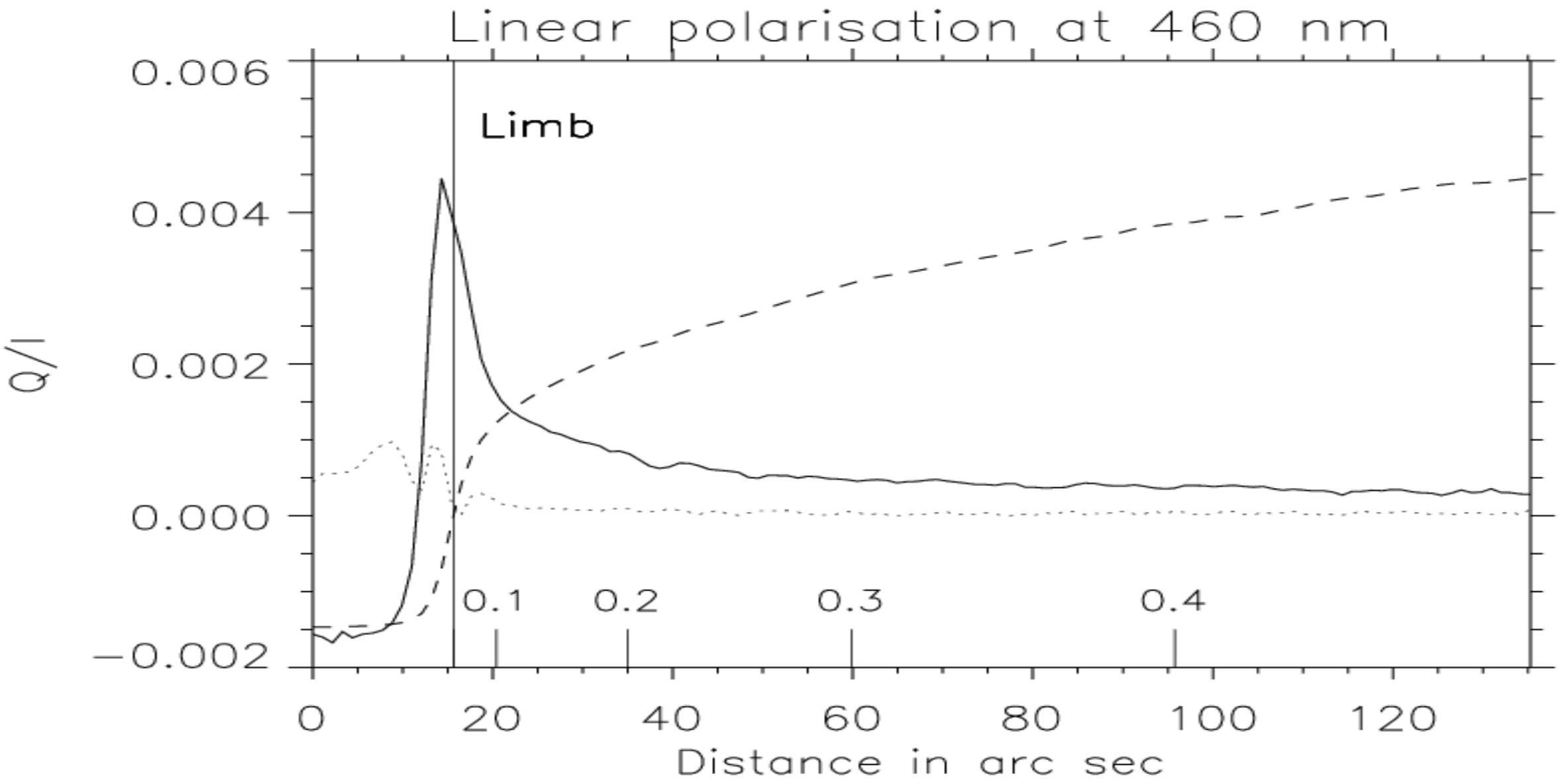

**Figure 16** : *Linear polarization (Q/I) at 4600 Å (bandwidth 100 Å, solid line) as a function of limb distance (arcsec or $\mu = \cos(\theta)$). Dashed line : intensity profile. Dotted line: error bar. Courtesy Paris observatory.*

## 5 – POLARIMETRY WITH THE BEAM SPLITTER SHIFTER

The present version of our single beam polarimeter has some limitations which can be removed with a dual beam. Polarization signals I + P, I - P (P = Q, U, V) are observed sequentially with a simple dichroïc linear polarizer. A dual beam is available since 2006 (figure 17), consisting of a birefringent linear polarizing beam splitter (spath) which allows to observe simultaneously the two linear combinations, so that I + P and I – P are focused at the same time on the CCD detector. As a consequence, the field of view (slit length) is reduced by a factor two, from 140" to 70" along the sun. The beam splitter includes a multi-slit and is located at focus F2 (figure 18) just at the entrance of the spectrograph. This improvement is of special importance for the polarimetry of solar structures observed on the disk (such as active regions, flux tubes) or at the limb (spicules, prominences) which require perfect coalignment and simultaneity of I + P and I - P signals, in order to eliminate seeing induced polarimetric cross talk. But observations of the second solar spectrum does not suffer from this default, because of statistical effects due to the large amount (thousands) of spectra required to achieve the convenient signal to noise ratio. Beam exchange is available by voltage adjustment of the two retarders, this is usefull for the measure of weak polarization signals, in order to observe I + P and I – P at time t, then I – P and I + P at time t + Δt. The beam exchange is the most precise method for deep polarimetry of the second solar spectrum, because it does not require any flatfield.

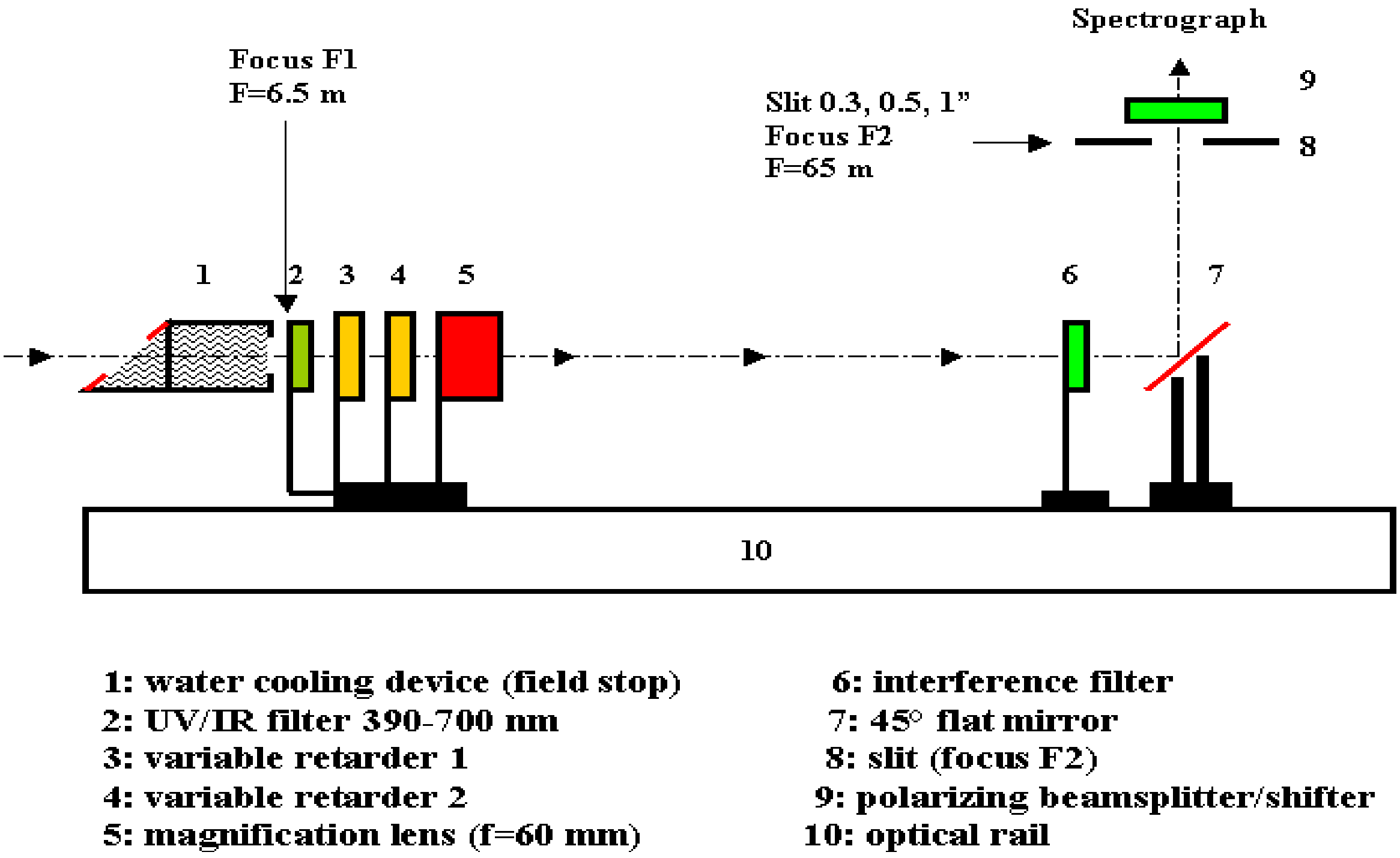

**Figure 17** : *The experiment setup with the polarizing beam splitter (9). Courtesy Paris observatory.*

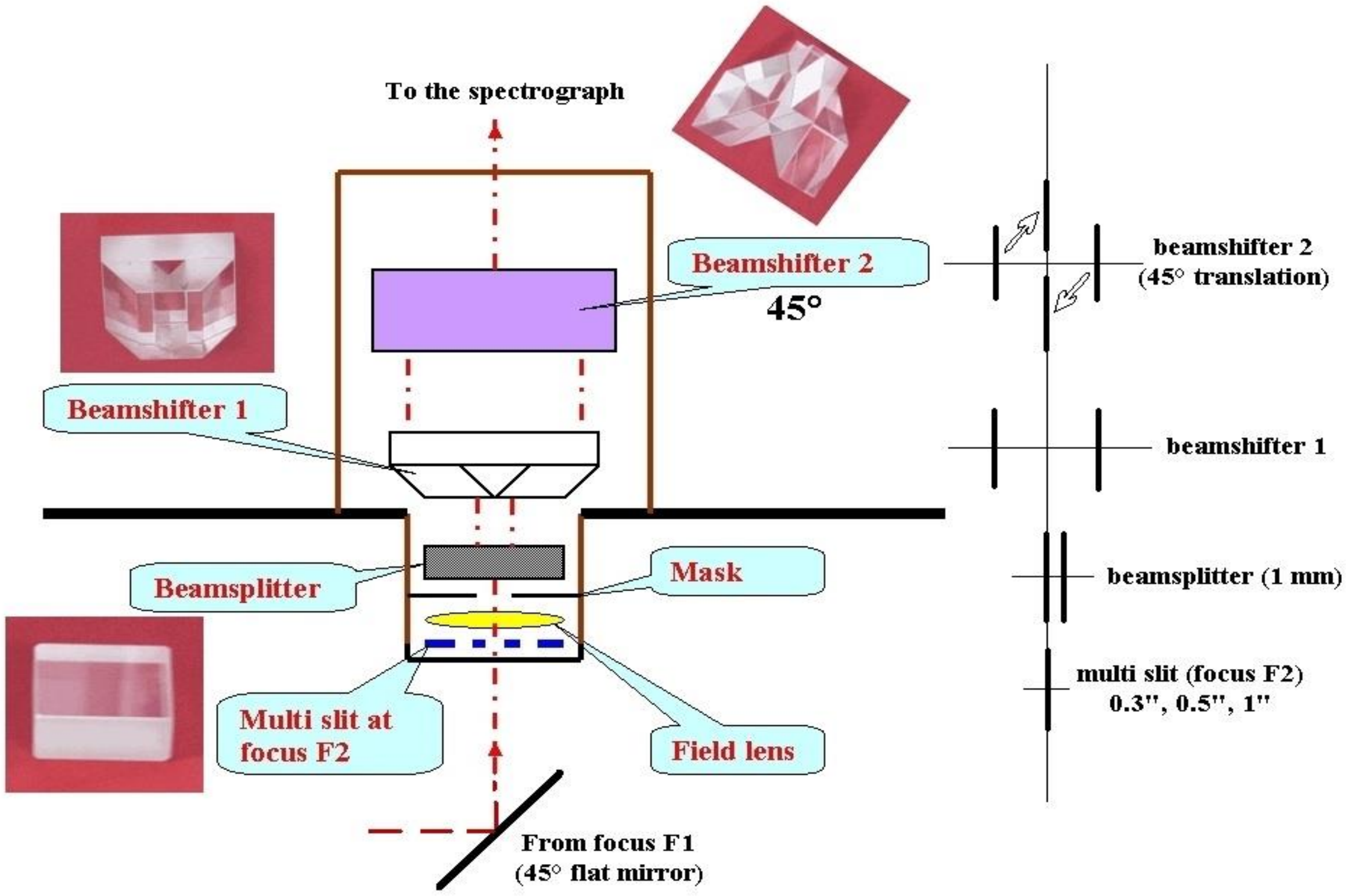

**Figure 18** : *Details of the polarizing beam splitter shifter designed by Meir Semel. The beamsplitter is made of birefringent spath of 10 mm thickness and delivers two images of the entrance slit in two orthogonal polarizations states (1 mm translation). Beamshifter 1 increases the distance between images, while beamshifter 2 realigns the two images of the slit for injection into the spectrograph. The length of the slit is 70" and the width is either 0.3", 0.5" or 1". Courtesy Paris observatory.*

## 6 – NEW RESULTS OF CAII K SPECTRO-POLARIMETRY

Many observations of the CaII K line (3933.7 Å) were performed in polarimetric mode, in imagery mode near the limb (for the continuum polarization) and in spectroscopy for the measure of the Hanle depolarization in the line core (K3), whih allows to estimate the unresolved turbulent magnetic field. Some observations of the circular polarization above susnpots (Zeeman effect) were also performed.

### Linear polarization of the CaII K continuum at the limb (imagery mode)

We studied the linear polarization of the pseudo continuum at 393 nm (10 nm bandpass) using the setup of figure 14 and the results are shown in figures 19 and 20. The polarization rate Q/I increases towards the limb where it reaches 0.6%. Thousands of couples I + Q and I – Q were observed alternatively. The data processing includes limb detection before summing I + Q and I – Q images in order to improve the signal to noise ratio. The zero polarization level is set by the flat field at the disk center where the linear polarization rate is supposed to vanish.

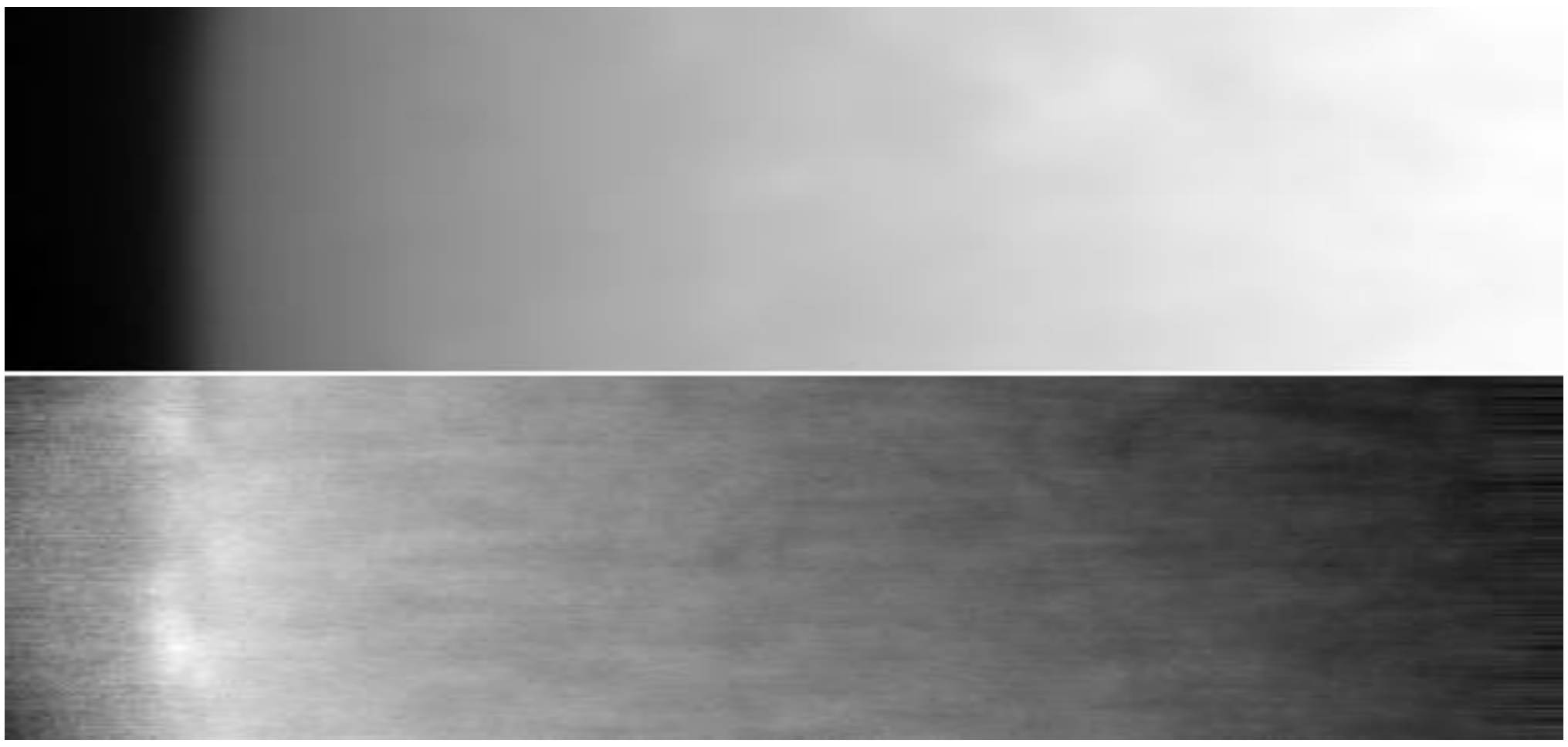

**Figure 19** : *Polarization of the continuum at 393 nm through a 10 nm filter in imagery mode. Top: intensity I. Bottom: polarization rate Q/I. Courtsey Paris observatory.*

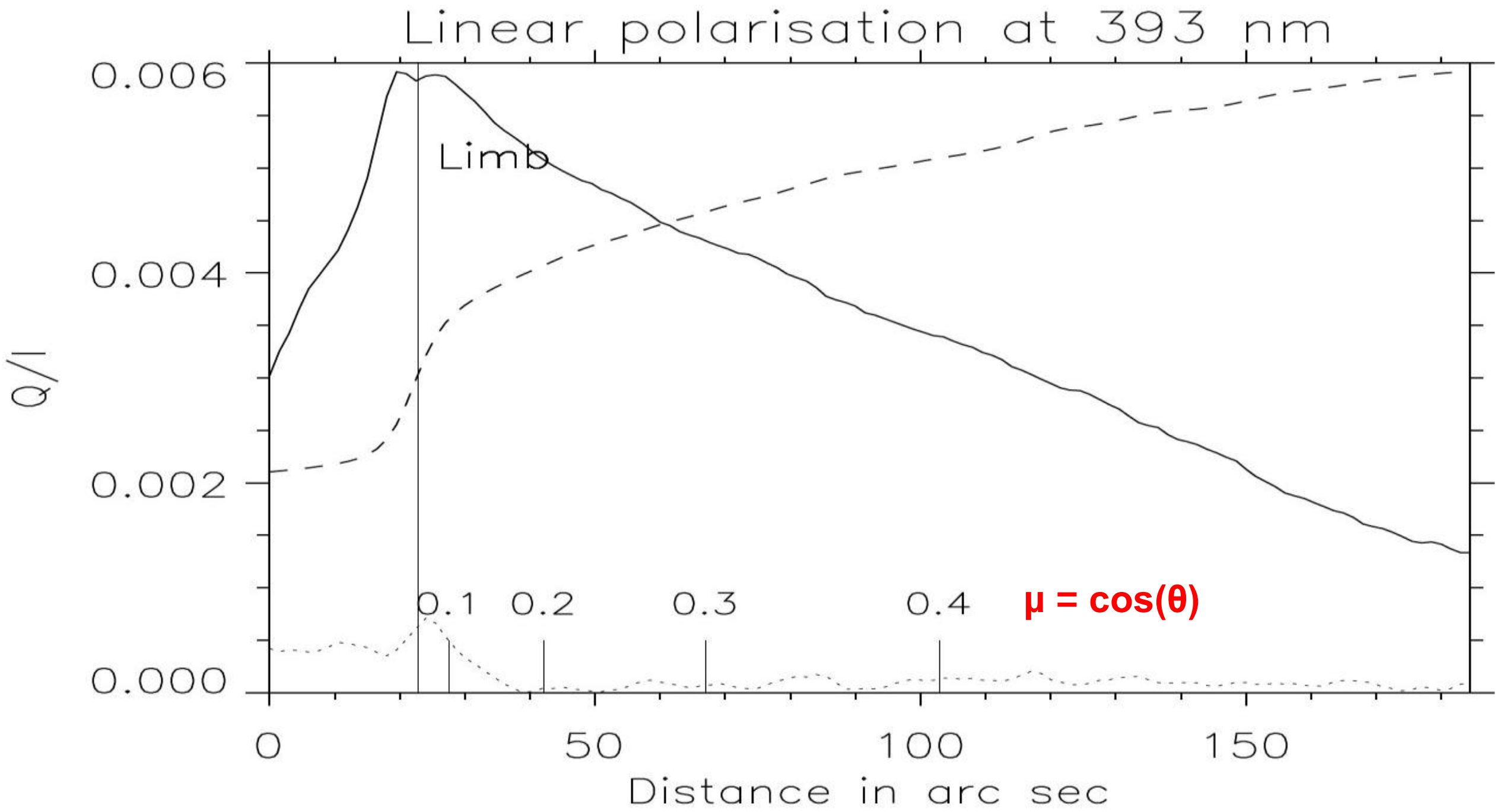

**Figure 20** : *Polarization rate Q/I of the continuum at 393 nm through a 10 nm filter, as a function of limb distance (μ = cos(θ) varies from 0 to 0.5). The dashed line is the intensity I of the limb profile and the dotted line is the error bar. Courtsey Paris observatory.*

## Circular polarization of the CaII K line above sunspots (spectroscopic mode)

Some observations of the circular polarization (movie 2) were done above an active region using the setup of figure 6 in spectroscopic mode (figure 21). In sunspots, the line core (K3) is in emission (in comparison to the quiet Sun, see Kneer & Mattig, 1978 ; Mattig & Kneer, 1978) ; K3 is sentisive to the Zeeman effect (g = 1.17) and was investigated by Martinez Pillet *et al* (1990). A cross section of the chromosphere above the sunspot is displayed in figure 22. It shows that the V/I signal, as a function of wavelength, is proportional to the quantity (1/I) (dI/dλ), which means the the weak field approximation applies. Hence, we deduce the line of sight magnetic field B// from the usual formula:

$V/I(\lambda) = 4.67\ 10^{-13}\ B_{//}\ g\ \lambda^2\ (1/I)\ (dI/d\lambda)$,

which gives B// = 580 G. This is the magnetic field in the chromosphere above sunpsots, while other lines in the vicinity of CaII K (such as iron lines) form in the photosphere and remain in absorption in sunspots, explaining the opposite Stokes V pattern.

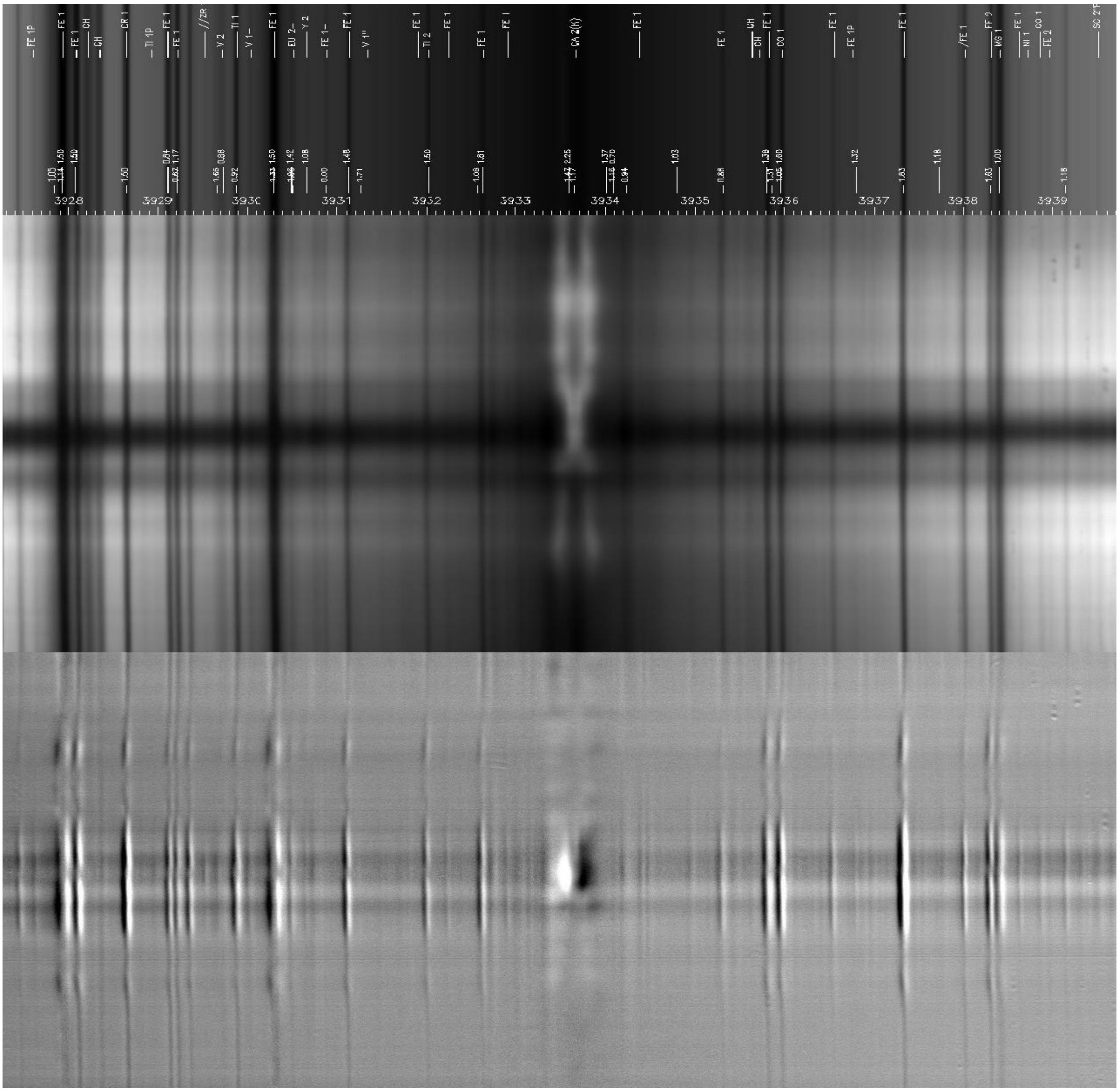

**Figure 21** : *Circular polarization in sunspots. Top: atlas with wavelength calibration in abscissa, spectral line identification (top labels) and Landé factors (bottom labels). Centre: intensity I(λ, x) above an active region and sunspot (umbra and penumbra). Bottom: circular polarization rate V/I(λ, x). Courtesy Paris observatory.*

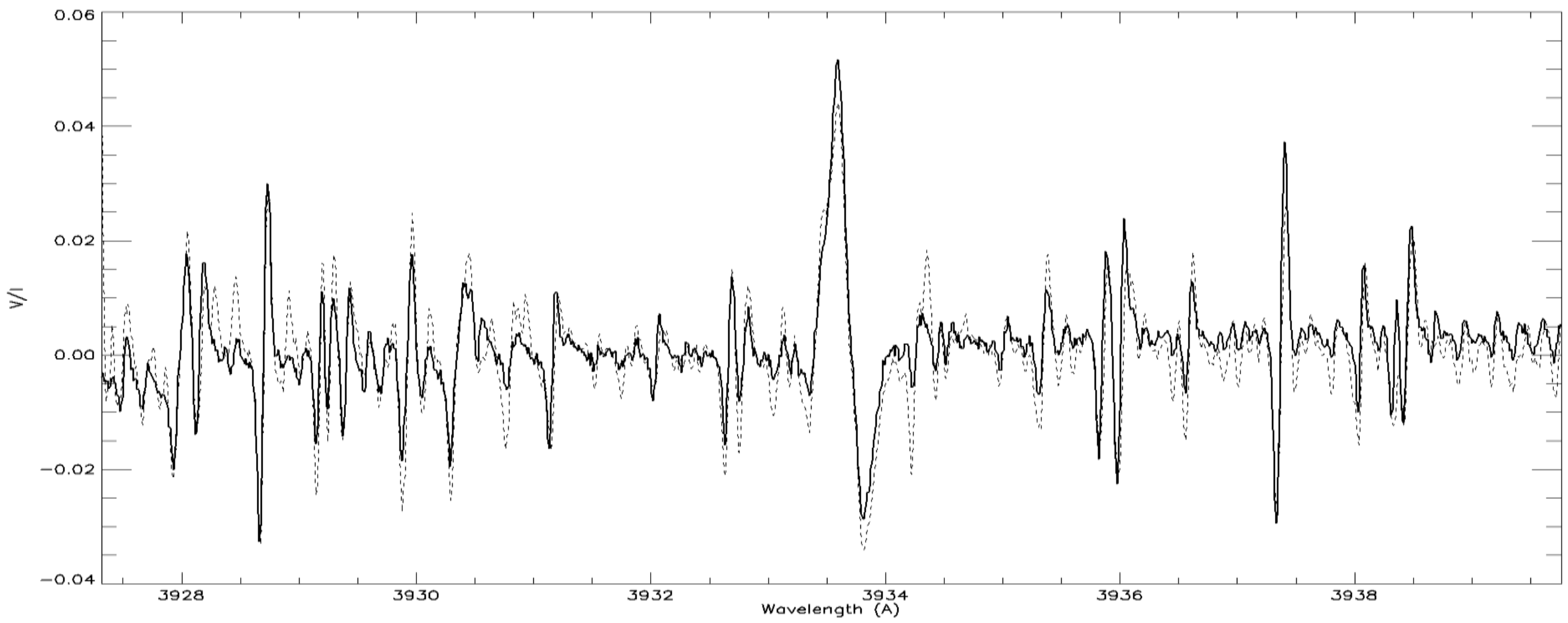

**Figure 22** : *Photosphere and chrompsphere above the sunspot umbra. Solid line: circular polarization rate V/I(λ). Dotted line: (1/I) dI/dλ showing that the weak field approximation applies (wavelength in abscissa, CaII K3 exhibits V/I = 0.05). Courtesy Paris observatory.*

### Linear polarization of the CaII K line at the limb (spectroscopic mode)

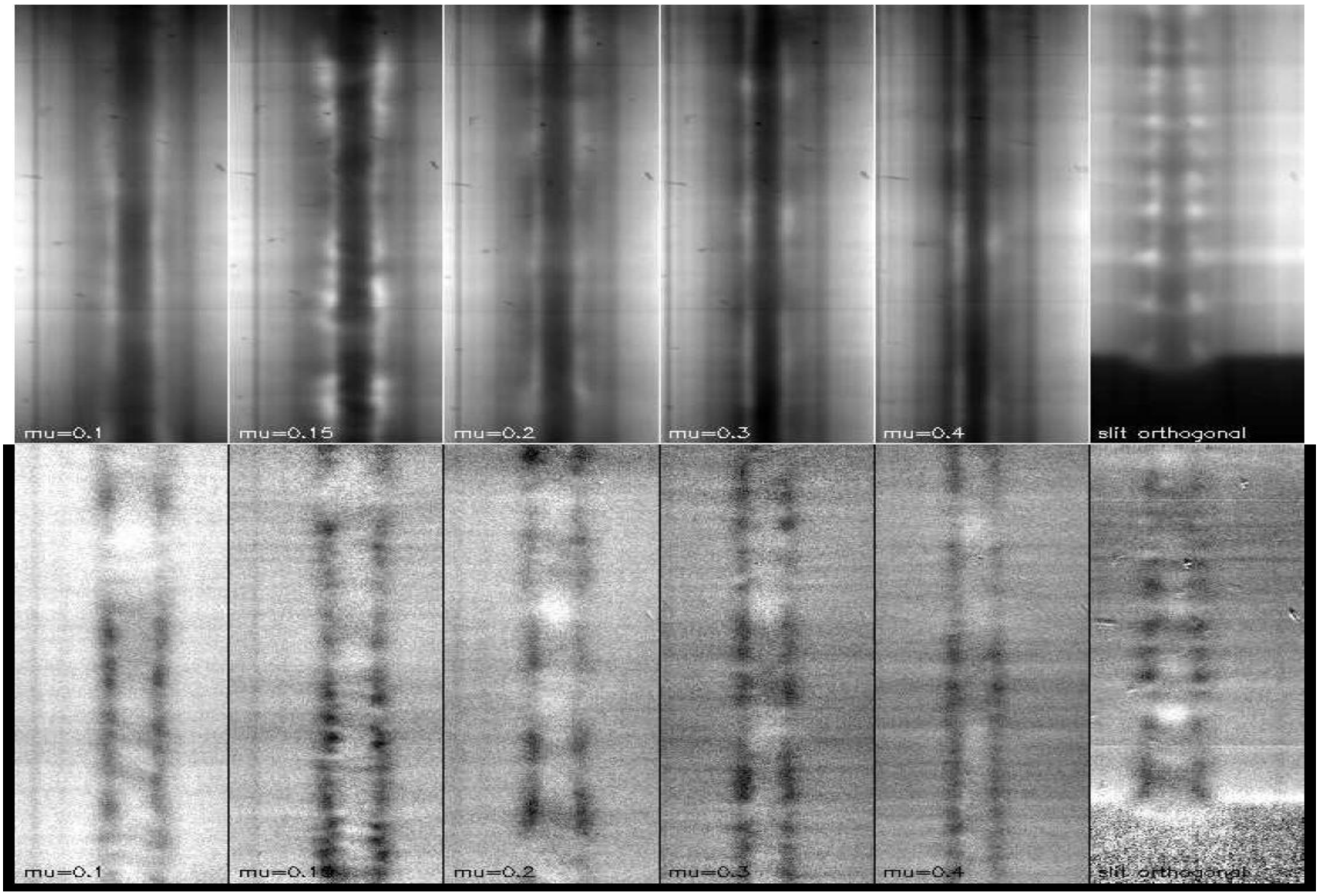

**Figure 23** : *Top: CaII K intensity spectra I(λ,x) with slit parallel to the limb (μ = 0.1, 0.15, 0.2, 0.3, 0.4) and slit orthogonal to the limb . Bottom: Polarization rate Q/I(λ,x). Wavelength (λ) in abscissa. Slit direction (x) in ordinates. Courtesy Paris observatory.*

Observations of the linear polarization of CaII K were performed with the setup of figure 6 in spectroscopic mode (figure 23). The slit was either parallel to the limb or orthogonal to the limb. When parallel to the limb, the polarization was summed along the 140" slit in order to increase the signal to noise (S/N) ratio (integration over 700 pixels increasing the S/N by 25), leading to the results of figures 24 and 25 which show the line intensity I(λ) and the polarization rate Q/I(λ) for various μ values (0.1, 0.14, 0.2, 0.28, 0.4) corresponding to limb distances of 5”, 10”, 20”, 40” and 80”. At line centre (K3), our results clearly show a depolarization in comparison to the scattering polarization rate predicted by Holzreuter *at al* (2006) and Holzreuter (2009) as a function of μ, using atmospheric models such as the FAL A, C or X. The presence of unresolved turbulent magnetic fields may explain this depolarization through the Hanle effect. Figure 26 shows the magnetic field computed using the method described by Stenflo (1982). We measured the depolarization with respect to the prediction given by Holzreuter (2009) with FAL C and X. It can be explained by a magnetic strength in the range 15-22 G. It has the tendancy to increase towards the limb, but we think that this behaviour is not significant because in the error bar of about 5 G.

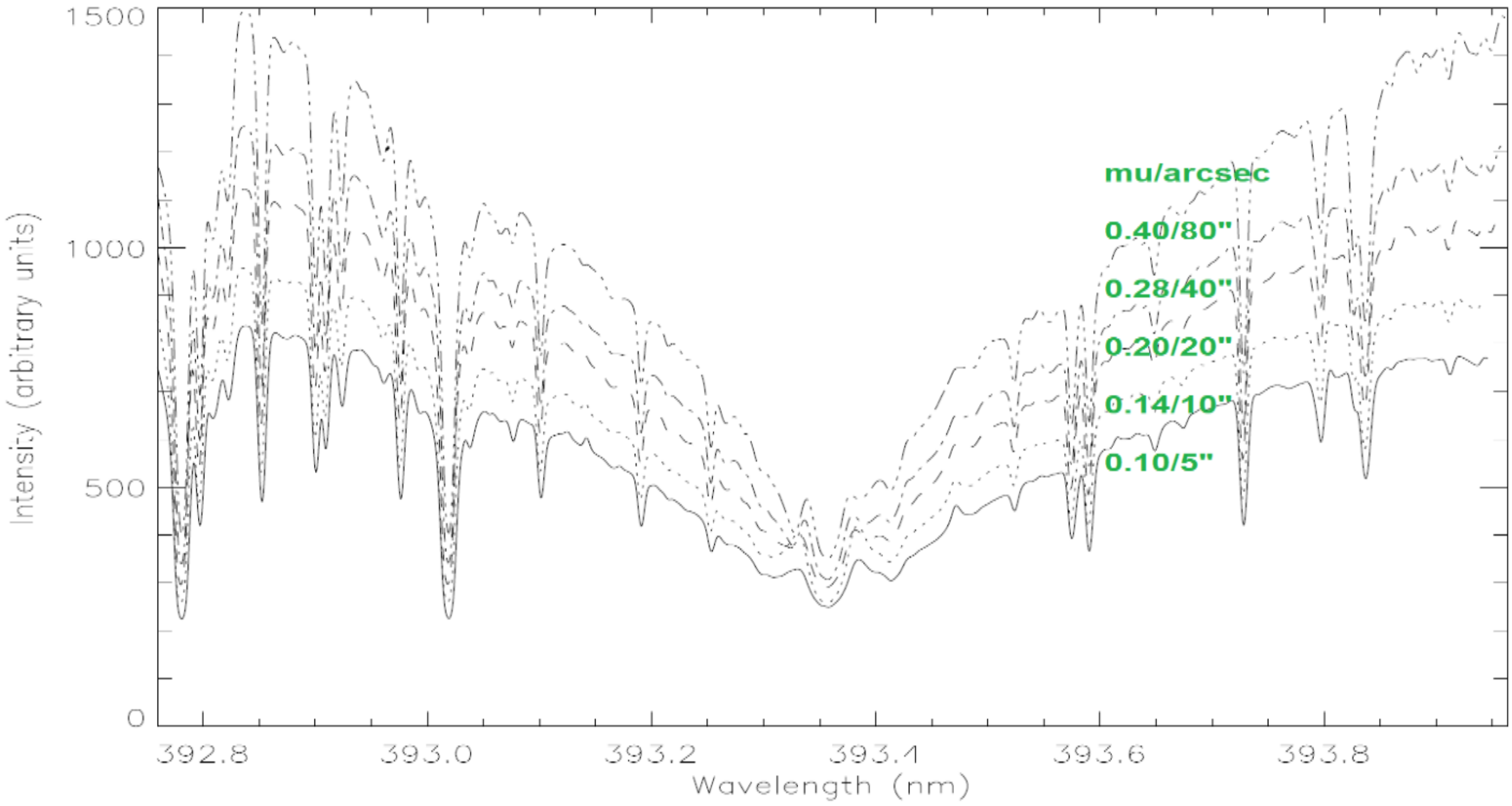

**Figure 24** : *CaII K line profiles at various limb distances (μ = 0.10, 0.14, 0.20, 0.28, 0.40 corresponding to 5”, 10”, 20”, 40”, 80”). The slit is parallel to the limb. Data are averaged along the slit. Courtesy Paris observatory.*

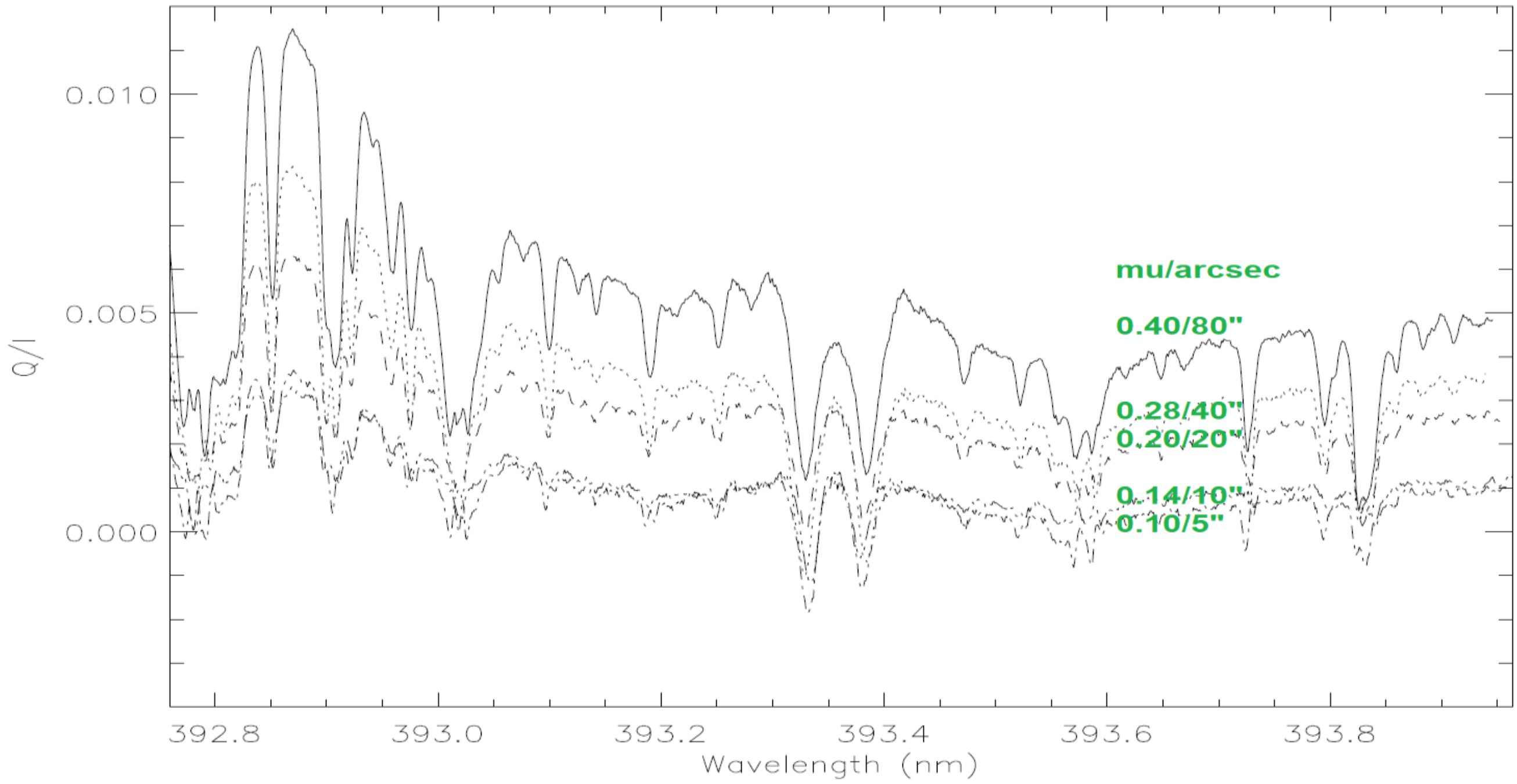

**Figure 25** : *CaII K linear polarization profiles Q/I at various limb distances (μ = 0.10, 0.14, 0.20, 0.28, 0.40 corresponding to 5”, 10”, 20”, 40”, 80”). The slit is parallel to the limb. Data are averaged along the 140” slit. Courtesy Paris observatory.*

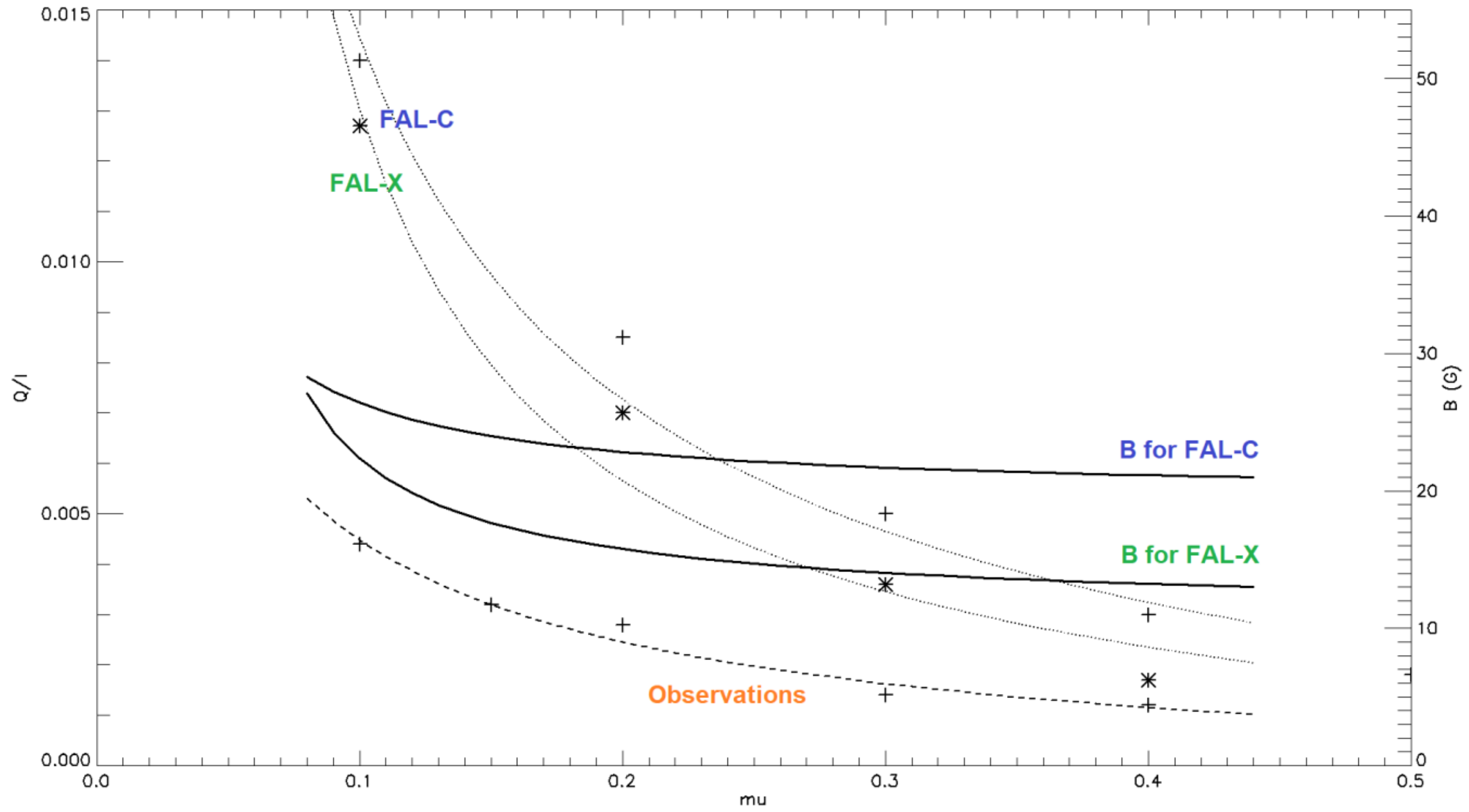

**Figure 26** : *Dashed line: observed linear polarization rate Q/I as a function of μ. Thin lines: maximum scattering polarization for FAC C and FAL X models (after Holzreuter, 2009) with no magnetic field. Thick lines: turbulent magnetic field deduced from the Hanle depolarization using the method of Stenflo (1982). Data curves are fitted by functions of the form a(1-μ²)/(μ+b) where a and b are constants determined by least squares. Courtesy Paris observatory.*

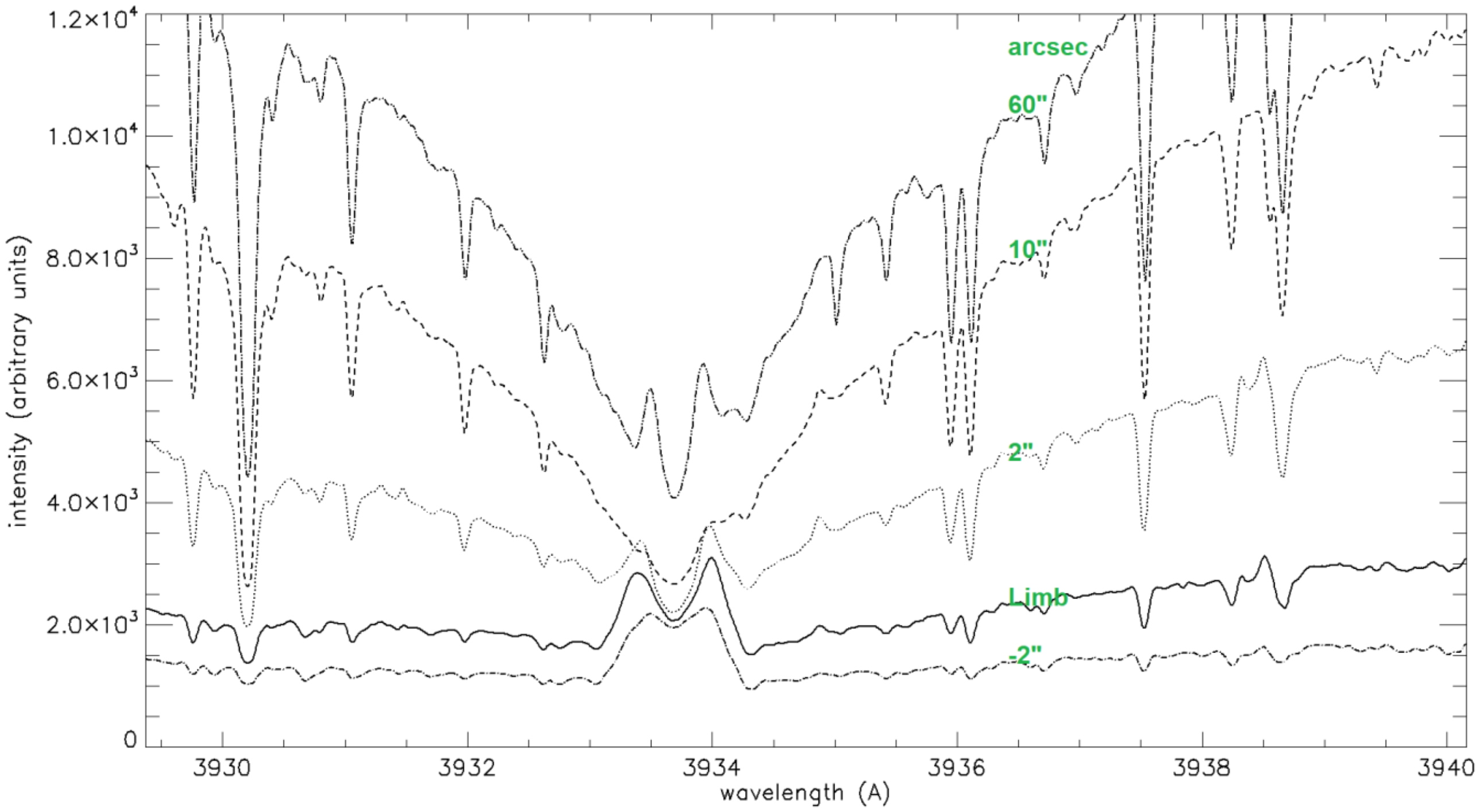

**Figure 27** : *CaII K line profiles at various limb distances (2" outside the solar disk, at the limb with line in emission, and 2", 10", 60" inside the disk). The slit is orthogonal to the limb. Data are averaged over 1" only and are more noisy than those of figure 24. Courtesy Paris observatory.*

Figures 27 and 28 show observations with the slit orthogonal to the limb. This strategy has the advantage of crossing the limb (movie 3); on the contrary, it is not possible to integrate data along the slit, so that several thousands of couples I + Q and I – Q must be recorded instead of one hundred to get the appropriate signal to noise ratio. This is a too long process (hours in the case of CaII K which requires long exposure time) to guarantee a perfect stability of the spectrograph (some parasitic shifts in wavelength and fluctuations of the slit position occur and cannot be fully corrected). We just provide the results (intensity profiles and polarization rate Q/I at the limb, at 2" outside the Sun and at 2", 10", 60" inside on the disk) for information, but we are more confident concerning polarization results of figure 25 with the slit parallel to the limb, allowing much faster observations and stable spectroscopic conditions.

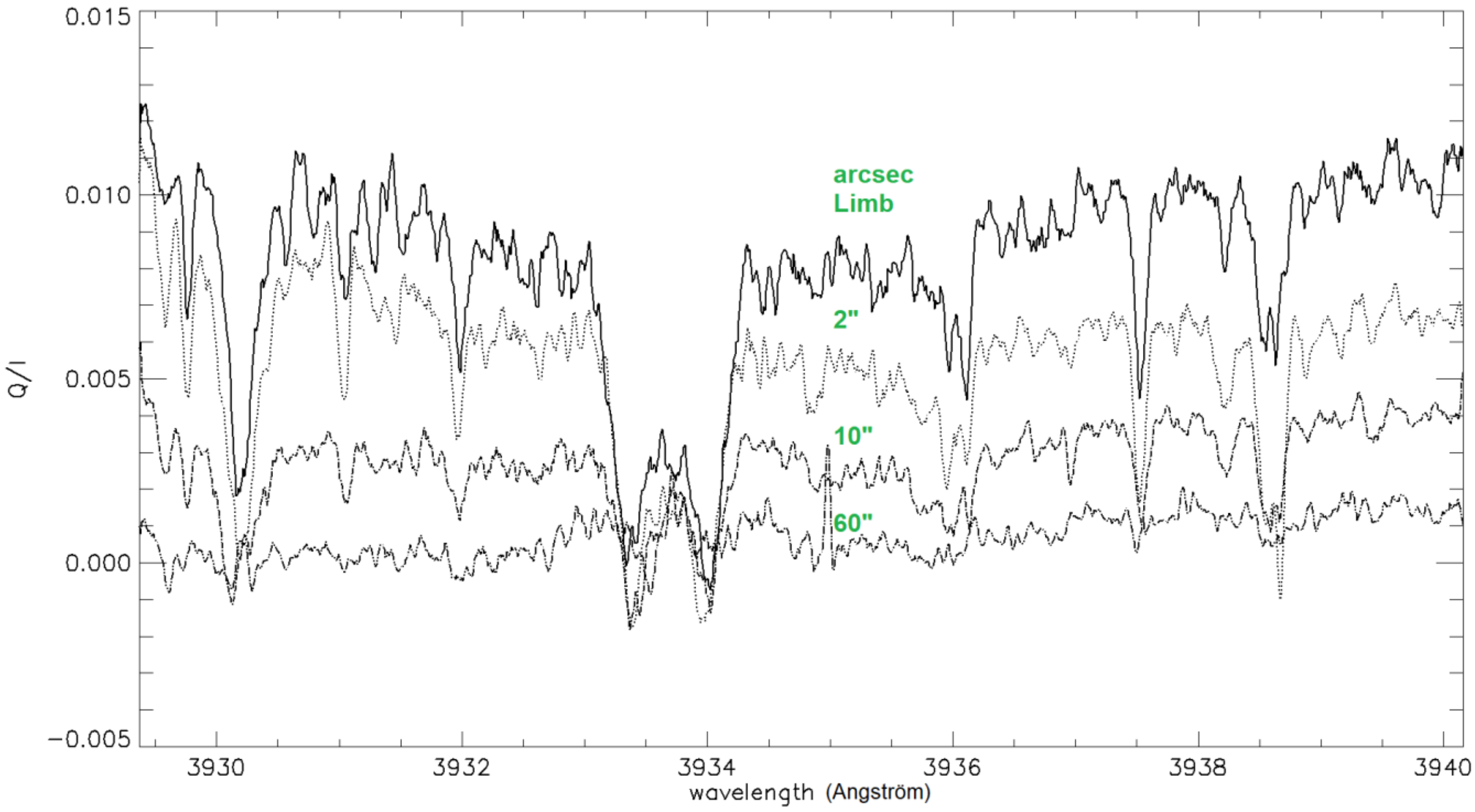

**Figure 28** : *CaII K linear polarization profiles Q/I at various limb distances (limb, 2", 10", 60"). The slit is orthogonal to the limb, so that data are averaged over 1" only and are more noisy than those of figure 25. Courtesy Paris observatory.*

## 7 – CONCLUSION

We described in this paper the evolution of the polarimeters operating at the Pic du Midi Turret Dome, in imagery, spectroscopy or imaging spectroscopy modes. They are based on Liquid Crystal Variable Retarders are are convenient for the analysis of the Zeeman effect in active regions or scattering polarization (and Hanle depolarization in the presence of unresolved turbulent magnetic fields). We also showed new results concerning the CaII K circular polarization in sunspots or linear polarization near the limb and deduced magnetic field strength in the range 15-22 G from the Hanle depolarization.

## 8 - REFERENCES

## 9 - ON-LINE MATERIAL (MPEG4 MOVIES)

Movie 1 : Granulation seen in NaD1, imaging spectroscopy (MSDP mode) at two altitudes, in the photosphere (line wings) and in the low chromosphere (line core).

https://drive.google.com/file/d/1-uTJ9Q1zUDU5gxczC1NrpU-17UFgDloD/view?usp=share_link

Movie 2 : Circular polarization in an active region sunpot, CaII K line (wavelength in abscissa, direction of the slit in ordinates), alternatively I+V and I-V polarization signals.

https://drive.google.com/file/d/1r1O0hD-pQWF_xLAZXti9-5KS-vgwccpF/view?usp=share_link

Movie 3 : CaII K line profile at various distances (arc sec) from the limb, from 75” (inside the disk) to -5” (outside the disk)

https://drive.google.com/file/d/1MGsh9SlaNIul7yxHMX8o0C1-bqNT5VrC/view?usp=share_link